\documentclass[english,twocolumn,english,aip,reprint]{revtex4-1}
\usepackage[T1]{fontenc}
\usepackage[latin9]{inputenc}
\usepackage{xcolor}
\usepackage{babel}
\usepackage{array}
\usepackage{verbatim}
\usepackage{booktabs}
\usepackage{textcomp}
\usepackage{amsmath}
\usepackage{graphicx}
\usepackage[unicode=true,pdfusetitle,
 bookmarks=true,bookmarksnumbered=false,bookmarksopen=false,
 breaklinks=false,pdfborder={0 0 1},backref=false,colorlinks=true]
 {hyperref}
\hypersetup{
 urlcolor=blue,linkcolor=red,citecolor=blue}

\makeatletter

\providecommand{\tabularnewline}{\\}

\usepackage{setspace}

\makeatother

\begin{document}
\title{Plasma-assisted molecular beam epitaxy of SnO(001) films: Metastability,
hole transport properties, Seebeck coefficient, and effective hole
mass}
\author{Melanie Budde}
\affiliation{Paul-Drude-Institut f\"ur Festk\"orperelektronik, Leibniz-Institut im
Forschungsverbund Berlin e.V., Hausvogteiplatz 5-7, 10117 Berlin,
Germany}
\author{Piero Mazzolini}
\affiliation{Paul-Drude-Institut f\"ur Festk\"orperelektronik, Leibniz-Institut im
Forschungsverbund Berlin e.V., Hausvogteiplatz 5-7, 10117 Berlin,
Germany}
\author{Johannes Feldl}
\affiliation{Paul-Drude-Institut f\"ur Festk\"orperelektronik, Leibniz-Institut im
Forschungsverbund Berlin e.V., Hausvogteiplatz 5-7, 10117 Berlin,
Germany}
\author{Christian Golz}
\affiliation{Department of Physics, Humboldt-Universit\"at zu Berlin, Newtonstrasse
15, 12439 Berlin,Germany}
\author{Takahiro Nagata}
\affiliation{National Institute for Materials Science (NIMS), Tsukuba, Ibaraki
305-0044, Japan}
\author{Shigenori Ueda}
\affiliation{National Institute for Materials Science (NIMS), Tsukuba, Ibaraki
305-0044, Japan}
\affiliation{Synchrotron X-ray Station at SPring-8, NIMS, Sayo, Hyogo 679-5148,
Japan}
\author{Georg Hoffmann}
\affiliation{Paul-Drude-Institut f\"ur Festk\"orperelektronik, Leibniz-Institut im
Forschungsverbund Berlin e.V., Hausvogteiplatz 5-7, 10117 Berlin,
Germany}
\author{Manfred Ramsteiner}
\affiliation{Paul-Drude-Institut f\"ur Festk\"orperelektronik, Leibniz-Institut im
Forschungsverbund Berlin e.V., Hausvogteiplatz 5-7, 10117 Berlin,
Germany}
\author{and Oliver Bierwagen}
\affiliation{Paul-Drude-Institut f\"ur Festk\"orperelektronik, Leibniz-Institut im
Forschungsverbund Berlin e.V., Hausvogteiplatz 5-7, 10117 Berlin,
Germany}
\author{}
\date{\today\\
}
\begin{abstract}
\noindent Transparent conducting or semiconducting oxides are an important
class of materials for (transparent) optoelectronic applications and
-- by virtue of their wide band gaps -- for power electronics. While
most of these oxides can be doped $n$-type only with room-temperature
electron mobilities on the order of 100\,cm$^{2}$/Vs, $p$-type
oxides are needed for the realization of $pn$-junction devices but
typically suffer from exessively low (<\textcompwordmark <1\,cm$^{2}$/Vs)
hole mobilities. Tin monoxide (SnO) is one of the few $p$-type oxides
with a higher hole mobility yet is currently lacking a well-established
understanding of its hole transport properties. Moreover, growth of
SnO is complicated by its metastability with respect to SnO$_{2}$
and Sn, requiring epitaxy for the realization of single crystalline
material typically required for high-end applications. Here, we give
a comprehensive account on the epitaxial growth of SnO, its (meta)stability,
and its thermoelectric transport properties in the context of the
present literature. Textured and single-crystalline, unintentionally-doped
$p$-type SnO(001) films are grown on Al$_{2}$O$_{3}$(00.1) and
Y$_{2}$O$_{3}$-stabilized ZrO$_{2}$(001), respectively, by plasma-assisted
molecular beam epitaxy and the epitaxial relations are determined.
The metastability of this semiconducting oxide is addressed theoretically
through an equilibrium phase diagram. Experimentally, the related
SnO growth window is rapidly determined by an in-situ growth kinetics
study as function of Sn-to-O-plasma flux ratio and growth temperature.
The presence of secondary Sn and SnO$_{x}$ $(1<x\le2)$ phases is
comprehensively studied by x-ray diffraction, Raman spectroscopy,
scanning electron microscopy, and x-ray photoelectron spectroscopy,
indicating the presence of Sn$_{3}$O$_{4}$ or Sn as major secondary
phases, as well as a fully oxidized SnO$_{2}$ film surface. The hole
transport properties, Seebeck coefficient, and density-of-states effective
mass are determined and critically discussed in the context of the
present literature on SnO, considering its strongly anisotropic effective
hole mass: Hall measurements of our films reveal room temperature
hole concentrations and mobilities  in the range of $2\cdot10^{18}$
to $10^{19}$\,cm$^{-3}$ and $1.0$ to $6.0$\,cm$^{2}$/Vs, respectively,
with consistently higher mobility in the single-crystalline films.
Temperature-dependent Hall measurements of the single-crystalline
films closest to stoichiometric, phase-pure SnO indicate non-degenerate
band transport by free holes (rather than hopping transport) with
dominant polar optical phonon scattering at room temperature. Taking
into account the impact of acceptor band formation and the apparent
activation of the hole concentration by 40--53\,meV, we assign tin
vacancies rather than their complexes with hydrogen as the unintentional
acceptor. The room temperature Seebeck coefficient in our films confirms
$p$-type conductivity by band transport. Its combination with the
hole concentration allows us to experimentally estimate the density
of states effective hole mass to be in the range of 1 to 8 times
the free electron mass.
\end{abstract}
\maketitle

\section{Introduction}

Most of the used semiconducting oxides are \textit{n}\nobreakdash-type,
reducing the applications of semiconducting oxides mainly to unipolar
devices. This is in part related to the low hole mobilities arising
from the strong localization of the O 2\textit{p} orbitals that make-up
the valence band maxima (VBM) with little dispersion, i.e. high hole
effective mass. For example, the hole mobility in the $p$-type semiconducting
oxides NiO:Li,\citep{Zhang2018} NiO,\citep{Karsthof2019} Cr$_{2}$O$_{3}$:Mg,\citep{Farrell2015}
and LaScO$_{3}$:Sr\citep{Zhang2015} is significantly lower than
1\,cm$^{2}$/Vs, being best described by polaronic hopping instead
of band transport.\citep{Zhang_pType,Karsthof2019} As a solution
hybridization between O 2\textit{p} and more spread orbitals by the
concept known as ``chemical modulation of the valence band'' has
been proposed to increase the dispersion of the VBM and thus decrease
the hole effective mass.\citep{Kawazoe1997}One candidate for hybridization
are lone-pair \textit{ns$^{2}$} orbitals.\citep{Zhang_pType} For
example, \textit{5s}$\mathrm{^{2}}$ in Sn$^{2+}$ of SnO forms a
stable configuration with the O 2\textit{p} orbitals, making SnO an
interesting material for $p$-type oxide electronics.\citep{Zhang_pType}
In fact, hole mobilities between 1 and 5~$\mathrm{\frac{cm^{2}}{Vs}}$
have typically been obtained by Hall measurements of SnO films.\citep{Zhang_pType,Wang2016,MBE_SnO_Schlom,Becker2019,SnO_Ogo_Transport}
More recently, hole mobilities as high as 30, 21, and 19\,~$\mathrm{\frac{cm^{2}}{Vs}}$
have been reported for polycrystalline SnO bulk ceramics,\citep{Miller2017}
optimized epitaxial SnO(001) layers,\citep{Minohara2019} and polycrystalline,
mixed SnO+Sn films,\citep{Record_mob_SnO} respectively. Thus, reasonably
high hole mobilities together with a direct bandgap absorption edge
around 2.6--3.2\,eV (and only weak optical absorption by its indirect
band gap of 0.6\,eV),\citep{Zhang_pType,Wang2016,MBE_SnO_Schlom}
fuel the interest in SnO as a \textit{p}-type semiconducting oxide
for transparent thin film transistor applications.\citep{SnO_Ogo_Transport}
The observed \textit{p}\nobreakdash-type conductivity of unintentionally
doped (UID) SnO has been correlated by first\nobreakdash-principle
calculations with Sn vacancies\citep{SnO_FirstPrinciple} or their
complexes with hydrogen\citep{Varley2013} acting as shallow acceptors,
whereas oxygen interstitials were predicted to be electrically inactive.\citep{SnO_FirstPrinciple,Varley2013}

SnO thin films have been grown by various methods, such as electron\nobreakdash-beam
evaporation\citep{SnO_Ebeam} or reactive DC magnetron sputtering,\citep{Record_mob_SnO}
both followed by thermal annealing, reactive ion beam sputter deposition,\citep{Becker2019}
pulsed laser deposition (PLD) from an oxide target\citep{SnO_Ogo_Transport,SnO_PLD,Minohara2019}
or a metallic Sn target\citep{Li2019}, and molecular beam epitaxy
(MBE).\citep{Hishita2010,MBE_SnO_Schlom,Nikiforov2020} The largest
challenge for the growth of phase pure SnO is its metastability with
respect to its stable relatives Sn and SnO$_{2}$.

At present the MBE growth of SnO is a rather unexplored field with
reports on the formation of polycrystalline SnO from the Sn-vapor
in the presence of pyrolyzed NO$_{2}$\citep{Hishita2010} or reactive
oxygen (followed by an annealing step)\citep{Nikiforov2020}. Using
MBE phase-pure, single crystalline SnO(001) films have so far only
been realized by subliming SnO$_{2}$ source material onto the heated
r-plane sapphire substrate without supplying additional oxygen.\citep{MBE_SnO_Schlom}
This is related to the fact that sublimation of SnO$_{2}$ produces
gaseous SnO and oxygen species.\citep{Hoffmann2020} 

In this study, we demonstrate the growth of textured and single-crystalline
SnO(001) films on Al$_{2}$O$_{3}$(00.1) and Y$_{2}$O$_{3}$-stabilized
ZrO$_{2}$(001) {[}YSZ(001){]}, respectively, from the Sn-vapor using
oxygen plasma-assisted MBE. After discussing the temperature-composition
phase diagram of the Sn-O-system obtained from thermochemical considerations,
we experimentally establish the related growth window using in-situ
analytics of the growth rate of SnO$_{2}$ and the desorption of SnO
from the growth front. The formed phase(s), epitaxial relation to
the substrate, and structural properties of films grown at different
conditions are shown. The hole transport properties are determined
and discussed in the context of existing literature on SnO. In particular,
the room-temperature hole concentration and Seebeck coefficient of
all films are determined and utilized to estimate the density-of-states
effective hole mass. Temperature-dependent hole transport properties
reveal band transport with a hole mobility limited by polar optical
phonon scattering and a free-hole activation energy of 53\,meV which
is compared to theoretically predicted acceptor ionization energies.

\section{Experiment}

SnO$_{x}$ films were grown by plasma assisted molecular beam epitaxy
as generally described in Ref.~\citenum{Vogt2015} on 2\nobreakdash-inch
c\nobreakdash-plane sapphire {[}Al$_{\mathrm{2}}$O$\mathrm{_{3}}$(00.1){]}
or quarters of 2\nobreakdash-inch YSZ(001) substrates. Both types
of substrates were covered with 1~\textmu m titanium by sputter deposition
on the rough backside to improve its radiative heating from the SiC
heating filament. The growth temperature $T_{g}$ was measured by
a thermocouple between substrate and heating filament. To improve
heating and layer uniformity the substrate was continuously rotated
at two rotations per minute. Sn (7N purity) was evaporated from a
shuttered single filament effusion cell operated at 1175~$^{\circ}$C, resulting
in a beam equivalent pressure of $\approx\mathrm{1.2\cdot10^{-7}}$~mbar
 at the substrate position. Activated oxygen was provided by passing
a controlled flow of O$_{2}$ (6N purity) through a radio-frequency
(RF) plasma source (run at a fixed RF power of 300\,W) directed at
the substrate.  Before growth a 20--30\,min oxygen plasma cleaning
was performed at $T_{g}=$700~$^{\circ}$C and an oxygen flux of 0.5\,standard
cubic centimeters per minute (sccm). After that the substrate temperature
was ramped to the desired growth temperature and the oxygen flux was
reduced to the desired growth flux ($f_{\text{g}}$). Growth was initiated
and terminated by opening and closing the Sn-shutter, respectively.
After film growth the substrate temperature was ramped down at 0.5$^{\circ}$C/s
to 200\,$^{\circ}$C under the O-plasma (using the O-flux as during growth)
followed by further cooldown to room temperature in vacuum.

The growth rate and amount of desorbing $^{136}$SnO was measured
in-situ using laser reflectometry (LR) and line-of-sight quadrupole
mass spectrometry (QMS), respectively, as described in Ref.~\citenum{Vogt2015}.
These measurements allowed us to determine the growth window of SnO
on c\nobreakdash-plane sapphire substrates. Afterwards, individual
SnO$_{x}$ layers were grown on Al$_{\mathrm{2}}$O$\mathrm{_{3}}$(0001)
and YSZ(001) at slightly different growth conditions.

These SnO$_{x}$ layers were structurally investigated by X-ray diffraction
(XRD) and Raman spectroscopy. XRD was measured in a four-circle diffractometer
using Cu K$\alpha$-radiation and a 1\,mm detector slit. The Raman
spectroscopic measurements were performed at room temperature in the
backscattering geometry with optical excitation at wavelengths of
473 nm (photon energy of 2.61\,eV) by a solid-state laser and at
325 nm (3.81\,eV) by a He-Cd laser. The incident laser light was
focused by a microscope objective onto the sample surface. The backscattered
light was collected by the same objective, spectrally dispersed by
an 80-cm spectrograph and detected by a liquid-nitrogen-cooled charge-coupled
device. The Raman spectra were recorded in the polarized configuration
(parallel polarizations of incoming and scattered light) using a linear
polarizer to analyze the detected light. Top-view scanning electron
microscopy (SEM) images were taken from all films.

On a selected sample, qualitative depth profiling of the valence band
structure and Sn 3d$_{5/2}$ core level was performed by photoelectron
spectroscopy taking advantage of the dependence of the photoelectron
mean free path $\lambda=\lambda_{0}\times\cos\text{(TOA)}$ on the
take-off-angle (TOA, 0$^{\circ}$ corresponds to normal emission) and kinetic
energy of the photoelectron (equalling the difference of photon energy
and binding energy) using soft X-rays (1468.6eV, $\lambda_{0}\approx1.8$\,nm
in SnO, Thermo Sigma-probe XPS system, SXPES) for surface sensitivity
as well as hard X-rays (HAXPES, 5956.3\,eV, $\lambda_{0}\approx6.9$\,nm
in SnO, beamline BL15XU at the Spring-8 synchrotron\citep{Ueda2010})
for bulk sensitivity as described in Ref.\,\citenum{Nagata2019a}.
Measurements were performed using the TOAs of 9.7, 54.7, 84.7 and
5.0, 50.0, 70.0\,$^{\circ}$ in SXPES and HAXPES, respectively.

In addition, Seebeck and Hall measurements in the van\nobreakdash-der\nobreakdash-Pauw
geometry, as described in Ref.\,\citenum{Preissler2013}, were used
to investigate the transport properties of the different layers at
room temperature. Temperature dependent Hall measurements on a selected
SnO sample were performed as described in Ref.\,\citenum{Golz2019}.

The thermal stability of the SnO phase was investigated using rapid
thermal annealing (RTA) at diffrent tempratures in nitrogen, oxygen
and forming gas ($\mathrm{N_{2}+H_{2}}$) at atmospheric pressure.
In addition, the stability under storage in air was investigated by
regular Hall measurements of a film over a period of 120 days.

\section{Thermodynamics of the growth window}

\begin{figure}
\begin{centering}
\includegraphics[width=8.5cm]{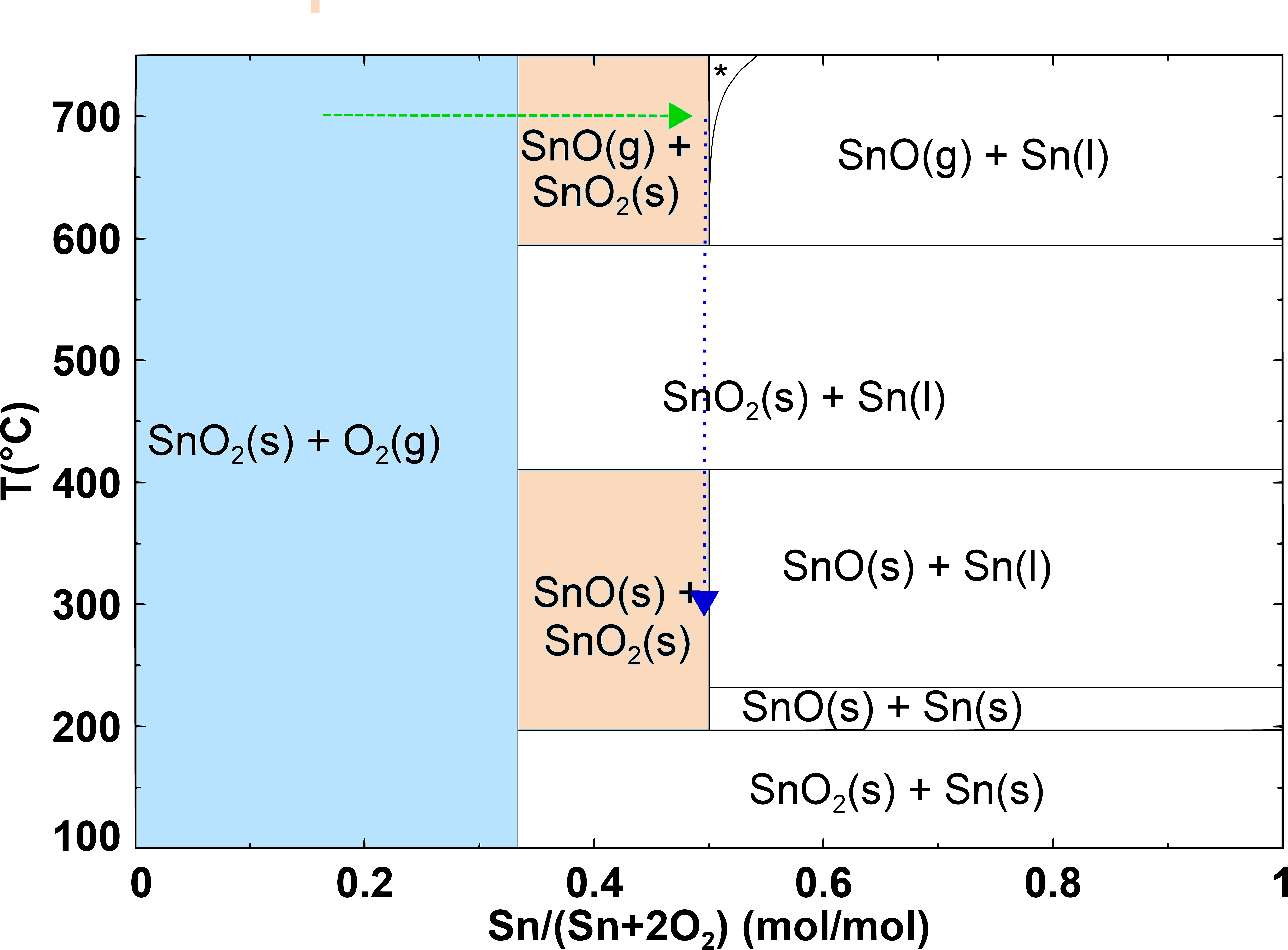}
\par\end{centering}
\caption{Equilibrium phase diagram of the Sn-O system as function of stoichiometry
and temperature at a typical pressure in the MBE growth chamber of
$10^{-6}$\,mbar. Stoichiometries of $n_{\text{Sn}}/(n_{\text{Sn}}+2n_{\text{\ensuremath{O_{2}}}})=0,\thinspace1/3,\thinspace1/2,$
and $1$ correspond to pure O$_{2}$, SnO$_{2}$, SnO, and Sn, respectively.
Constituents labeled ``(s)'', ``(l)'' and ``(g)'', are solid,
liquid, and gaseous, respectively. The phase labeled ``{*}'' denotes
SnO(g)\,+\,Sn(g). For comparison to Fig.\,\ref{fig:Oxygen-vs-Growth}
the light blue and light orange shaded areas corresponds to an oxygen-rich
and tin-rich growth regime of SnO$_{2}$, respectively, and the green
and blue arrows indicate the used strategy of determining the growth
window for SnO-growth as described in the following sub-section.\label{fig:phasediagram}}
\end{figure}
Common challenges for the growth of phase-pure SnO are its metastability
with respect to the disproportionation into SnO$_{2}$ and Sn, as
well as the adjustment of the stoichiometry to prevent the formation
of secondary SnO$_{2}$ or Sn phases. Fig.\,\ref{fig:phasediagram}
illustrates this situation with the help of the equilibrium Sn-O phase
diagram calculated by the FactSage 7.3 software package\citep{factsage}
as a function of stoichiometry $n_{\text{Sn}}/(n_{\text{Sn}}+2n_{\text{\ensuremath{O_{2}}}})$
and temperature. While solid phases {[}labeled by ``(s)''{]} of
Sn, SnO, and SnO$_{2}$ are present, the calcuations did not predict
stability of the intermediate oxides Sn$_{2}$O$_{3}$ and Sn$_{3}$O$_{4}$.
Note that despite the non-equilibrium nature of thin film growth,
equilibrium phase diagrams can provide guidance as discussed next:
Firstly, the stability region of SnO at temperatures between 197 and
410\,$^{\circ}$C (and disproportionation outside this region) rationalizes
why most SnO films have been obtained at growth or annealing temperatures
in this temperature range;\citep{SnO_PLD,Wang2016,Becker2019,Minohara2019,MBE_SnO_Schlom,Nikiforov2020}
secondly, during growth by reactive sputtering \citep{Becker2019,Record_mob_SnO},
PLD,\citep{SnO_Ogo_Transport,Li2019} or MBE\citep{Hishita2010} the
formation of secondary SnO$_{2}$ or Sn-phases has been controlled
by adjusting the stoichiometry of the source vapor, i.e., the oxygen
(background) pressure at fixed flux of SnO$_{x}$ from the source,
in qualitative agreement with the equilibrium stoichiometry dependence
in the phase diagram.  The blue or orange shaded regions as well
as colored arrows in Fig.\,\ref{fig:phasediagram} indicate the relation
to the in-situ determination of the growth window discussed next.

\section{Rapid in-situ kinetic determination of the growth window}

We have previously demonstrated that suboxide-possessing binary oxides
generally grow in MBE by a two\nobreakdash-step kinetics through
intermediate suboxide formation.\citep{Vogt_twoStep} For example,
during the growth of SnO$_{2}$, Sn is oxidized to the suboxide SnO
($\mathrm{Sn+O\rightarrow SnO}$) in the first step. In the second
step SnO is further oxidized to SnO$\mathrm{_{2}}$ ($\mathrm{SnO+O\rightarrow SnO_{2}}$)
if sufficient oxygen is available. Thus, at a sufficiently high Sn/O
flux ratio the second step is suppressed, preventing the growth of
SnO$_{2}$. In addition, a sufficiently high growth temperature leads
to desorption of the intermediate SnO that does not get further oxidized
and thus does not contribute to film growth.\citep{Tsai2009}

\begin{figure}
\begin{centering}
\includegraphics[width=8.5cm]{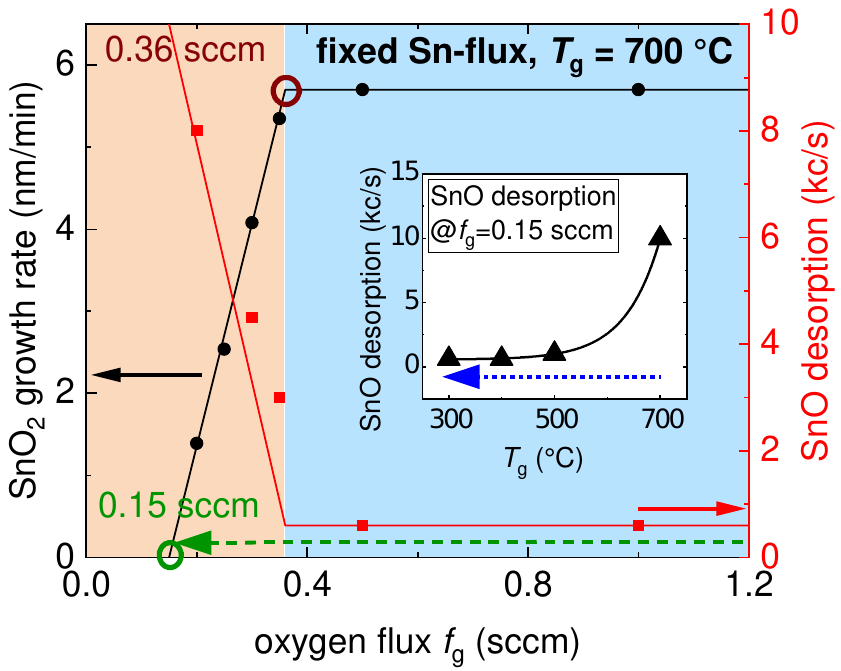}
\par\end{centering}
\caption{SnO$\mathrm{_{2}}$ growth rate (black discs) and desorbing SnO flux
(red squares) as function of oxygen flux, identifying two growth regimes:
oxygen rich (blue) and metal rich (orange). Red and black lines are
guides to the eye. The inset shows the desorbing SnO flux at fixed
oxygen flux as a function of growth temperature. To facilitate comparison
to the phase diagram the green dashed and blue dotted arrows correspond
to those in Fig.\,\ref{fig:phasediagram}.\label{fig:Oxygen-vs-Growth}}
\end{figure}
By taking advantage of this behavior we determined the stoichiometric
metal to oxygen flux ratio for the SnO formation by simultaneously
measuring the growth rate of SnO$\mathrm{_{2}}$ and the desorption
of SnO in-situ on a single Al$_{\mathrm{2}}$O$\mathrm{_{3}}$(0001)
substrate at $T_{g}=700$~$^{\circ}$C, which is high enough to result in desorption
of SnO that cannot get further oxidized. In contrast to Refs.\,\citenum{Tsai2009, Vogt2015},
in which the growth rates for varying Sn flux are discussed at fixed
O-flux, we kept the Sn-flux constant and gradually decreased the O-flux
from $f_{g}=1.0$\,to\,$0.2$\,sccm, which corresponds to a transition
from the SnO$_{2}$(s)\,+\,O$_{2}$(g) to the SnO(g)\,+\,SnO$_{2}$(s)
phase (green, dashed arrow in Fig.\,\ref{fig:phasediagram}). The
corresponding diagram shown in Fig~\ref{fig:Oxygen-vs-Growth} exhibits
a constant, high SnO$_{2}$ growth rate without SnO desorption for
$f_{g}$ ranging from $1$\,sccm to $0.36$\,sccm (blue shaded region).
This bevavior indicates oxygen-rich growth conditions under which
the entire Sn-flux is oxidized to SnO$_{2}$,\citep{Tsai2009,Vogt2015,Vogt_twoStep}
corresponding to a stoichiometry in the region of the SnO$_{2}$(s)\,+\,O$_{2}$(g)
phase in Fig.\,\ref{fig:phasediagram}. Decreasing $f_{g}$ from
$0.36$~sccm (orange shaded region) leads to a decreasing SnO$_{2}$
growth rate and simultaneously increasing SnO desorption, indicative
of Sn-rich growth conditions\citep{Tsai2009,Vogt2015,Vogt_twoStep}
and a corresponding stoichiometry in the region of the SnO(g)\,+\,SnO$_{2}$(s)
phase in Fig.\,\ref{fig:phasediagram}. In this case part of the
SnO which is formed in the first step desorbs due to an insufficient
O-flux for the second oxidation step.\citep{Vogt_twoStep} Consequently,
the extrapolated $f_{g}=0.15$\,sccm at which SnO$_{2}$ growth ceases
corresponds to the complete suppression of the second oxidation step,
an Sn/O-flux ratio for stoichiometric SnO formation, and the phase
boundary between SnO(g)\,+\,SnO$_{2}$(s) and SnO(g)\,+\,Sn(l)
at $n_{\text{Sn}}/(n_{\text{Sn}}+2n_{\text{\ensuremath{O_{2}}}})=0.5$
(marked by the end of the green arrow) in Fig.\,\ref{fig:phasediagram}.

Next, we determined the $T_{g}$-dimension of the SnO growth window
at the approximately stoichiometric oxygen flux of $0.15$\,sccm
for SnO formation by measuring the desorbing SnO-flux in-situ in the
range of 300\,$^{\circ}$C$\le T_{g}\le$700$^{\circ}$C on the same Al$_{\mathrm{2}}$O$\mathrm{_{3}}$(00.1)
substrate, corresponding to the blue, dotted arrow in Fig.\,\ref{fig:phasediagram}.
The results shown in the inset of Fig.\,\ref{fig:Oxygen-vs-Growth}
reveal negligible SnO desorption at $T_{G}\le500$\,$^{\circ}$C, indicating
the growth of solid SnO at the used fluxes.

\section{Epitaxial growth, phases, epitaxial relation}

Individual SnO$_{x}$ thin film samples were epitaxially grown on
YSZ(001) and Al$_{2}$O$_{3}$(00.1) within the delineated growth
window and the formed phases and structural properties were determined.
The following crystal structures and phases can (potentially) be found
in the grown samples: YSZ crystallizes in the cubic, fluorite structure
with a lattice parameter of about $a=0.512$\,nm, whereas sapphire
(Al$_{2}$O$_{3}$) possesses the corundum structure with lattice
parameters $a=0.4763$\,nm and $c=1.3003$\,nm. Sn crystallizes
in a tetragonal structure with lattice parameters $a=0.583$\,nm
and $c=0.318$\,nm.\citep{Deshpande1961} The best documented and
most stable SnO$_{x}$ phases are the tetragonal rutile SnO$_{2}$
with lattice parameters of $a=0.474$\,nm and $c=0.319$\,nm,\citep{Baur1956}
and tetragonal $\alpha$-SnO with $a=0.380$\,nm, $c=0.484$\,nm.\citep{Pannetier1980}
In addition, the intermediate phases Sn$_{2}$O$_{3}$ and Sn$_{3}$O$_{4}$
have been identified in the past both with monoclinic as well as triclinic
crystal structures. The monoclinic structure has been theoretically
predicted for both stoichiometries by the cluster expansion technique\citep{Seko2008}
and by ab initio calculations.\citep{Raman_Intermediate} In an early
experimental paper Lawson identified the triclinic structure for Sn$_{3}$O$_{4}$
using XRD,\citep{LAWSON1967} whereas later White et al. determined
Sn$_{3}$O$_{4}$ to be monoclinic by precession electron diffraction
measurements.\citep{White2010} For Sn$_{2}$O$_{3}$ a triclinic
phase has been identified using powder diffraction by Murken and Tr\"omel\citep{Murken1973}
and has been confirmed by Kuang et al.\citep{Kuang2016} for nanostructures
formed by a hydrothermal method. The $2\varTheta$- angles of XRD
reflexes of the triclinic phases of Sn$_{2}$O$_{3}$\citep{Murken1973}
and Sn$_{3}$O$_{4}$\citep{LAWSON1967} are similar enough to impede
an unambiguous distinction of both phases by XRD. Likewise, the spread
of theoretically and experimentally determined crystal structures
and lattice parameters for Sn$_{3}$O$_{4}$ and related $2\varTheta$-
angles of XRD reflexes make it difficult to unambiguously distinguish
Sn$_{3}$O$_{4}$ from other SnO$_{x}$ phases (including Sn) by XRD.

Raman spectroscopy is a powerful alternative method for phase identification.
In contrast to XRD, Raman spectra of monoclinic and triclinic Sn$_{3}$O$_{4}$
do not differ drastically.\citep{Raman_Intermediate,Berengue2010}
Furthermore, Eifert et al. have predicted distinctly different Raman
spectra for monoclinic Sn$_{3}$O$_{4}$ and Sn$_{2}$O$_{3}$ by
first principles calculations.\citep{Raman_Intermediate} In a series
of SnO$_{x}$ samples prepared by ion-beam sputtering with $x$ varying
from 1 to 2, they have only identified the presence of Sn$_{3}$O$_{4}$
as intermediate phase by Raman spectroscopy.\citep{Raman_Intermediate}
In agreement with that work we have not identified spectral features
which could be assigned to Sn$_{2}$O$_{3}$ in Raman spectra of our
films, either, and will therefore focus on Sn$_{3}$O$_{4}$ as intermediate
phase. The measured Raman spectra of all our epitaxial films are,
in general, well described by a superposition of contributions from
different oxide phases in addition to spectral features originating
from the substrate and metallic Sn in certain cases. The peak positions
of the dominant Raman peaks expected for the oxide phases SnO,\citep{Geurts1984,Song2015,Raman_Intermediate}
SnO$_{2}$,\citep{Raman_Intermediate,Katiyar1971,Lan2012}, and Sn$_{3}$O$_{4}$
\citep{Balgude2016,Raman_Intermediate,Song2015,Berengue2010} as well
as metallic Sn \citep{Olijnyk1992} are indicated in Figs.\,\ref{fig:SnOAl2O3}(c,d)
and \ref{fig:SnOYSZ}(c,d) as vertical dashed lines. In the case of
SnO$_{2}$ it has been taken into account that for the polarization
configuration of our experiments only the A$_{1g}$ phonon mode at
638 cm$^{-1}$ can be observed.\citep{Katiyar1971}

Whereas the optical probing depth $(\alpha_{i}+\alpha_{s})^{-1}$
in SnO is 260\,nm for excitation at 2.62\,eV, only 5 nm below the
surface is probed when excitation at 3.81 eV is chosen ($\alpha_{i}$
and $\alpha_{s}$ are the absorption coefficents at the photon energies
of the incoming and scattered light, respectively, taken from Ref.\,\citenum{MBE_SnO_Schlom}).
Consequently, the bulk properties of epitaxial films consisting mainly
of SnO can be investigated when using excitation at 2.62\,eV (similar
to the case of XRD measurements), while at 3.81\,eV just the near
surface layers are probed. Regarding the sensitivity for different
oxide phases, it has to be considered that the dielectric function
$\epsilon(\omega)$ of SnO\citep{MBE_SnO_Schlom} and the absorption
spectrum of Sn$_{3}$O$_{4}$\citep{Balgude2016} exhibit maxima in
the near ultraviolet spectral range which are connected with large
absolute values of the derivative $|d\epsilon/d\omega|$ both at photon
energies of 2.6 and 3.8 eV. Under this prerequisite, strong resonance
enhancements in efficiency of Raman scattering are expected.\citep{Cardona1982,Compaan1984}
For SnO$_{2}$, in contrast, the onset of strong optical absorption
at 4.28\,eV extracted from the ordinary diecelectric function does
not result in a particularly large value of $|d\epsilon/d\omega|$.\citep{Feneberg2014a}
As a consequence, the sensitivity for the detection of Raman signals
is expected to be better for SnO and Sn$_{3}$O$_{4}$ compared to
SnO$_{2}$ at both excitation energies (2.62 and 3.81\,eV) chosen
for our experiments.

\subsection{Growth on Al$_{2}$O$_{3}$(00.1) at different temperatures}

\noindent 

\noindent 
\begin{figure*}
\noindent \centering{}\includegraphics[width=15cm]{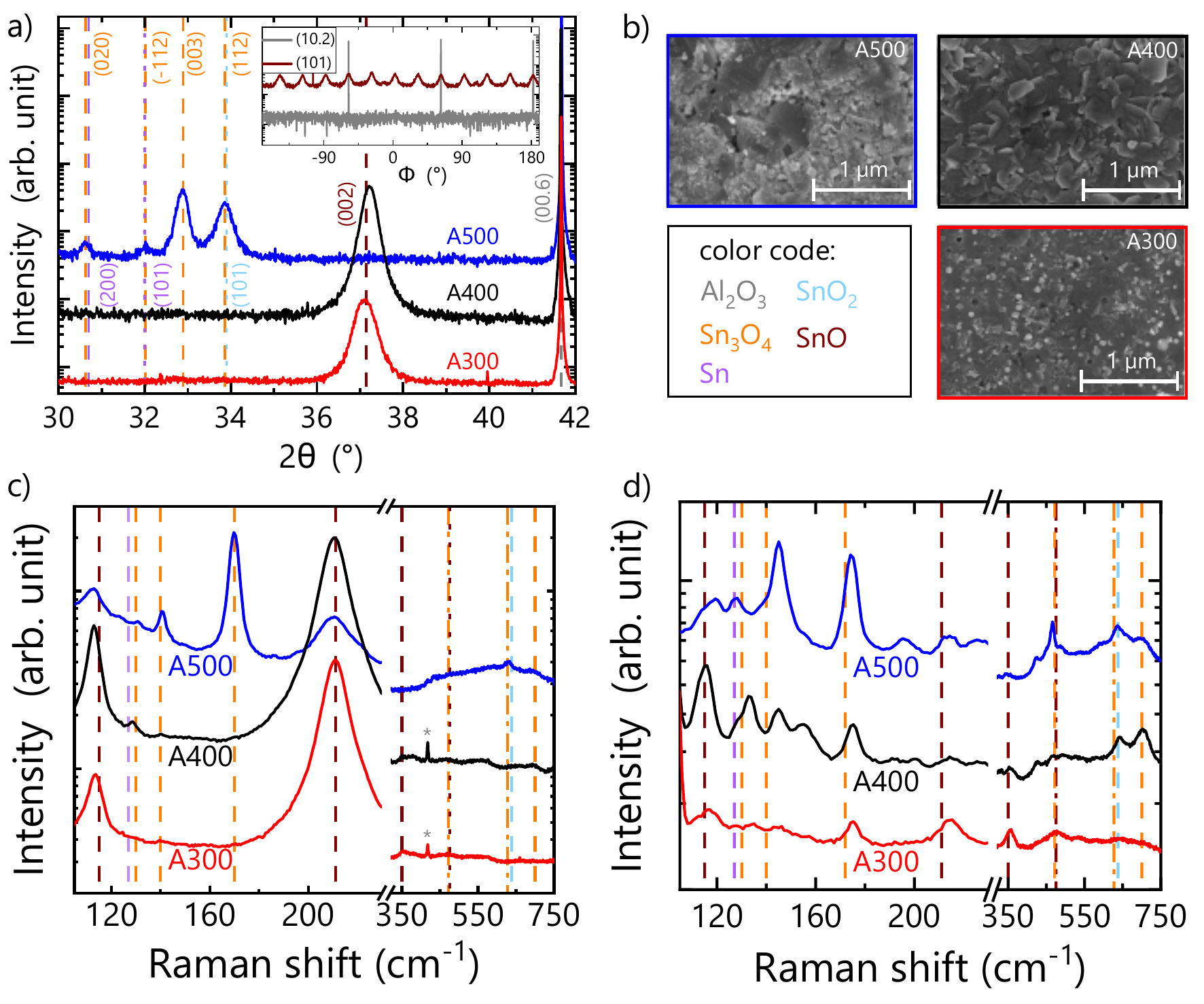}\caption{Structural characterization of SnO films grown on Al$_{2}$O$_{3}$(00.1)
at different substrate temperature. a) XRD symmetric out-of-plane
$2\Theta-\omega$ scan. Vertical lines indicate the expected positions
of reflexes of different phases as indicated by color code on the
right hand side. The inset shows a $\Phi$-scan of skew symmetric
substrate and SnO reflexes as indicated of A400. (b) Top view SEM
images. (c) Bulk-sensitive and (d) surface-sensitive Raman spectra
measured with an excitation wavelength of 473 and 325\,nm, respectively.
Vertical lines indicate the peak positions of dominant Raman active
phonon modes expected for different phases as indicated by the color
code in (b). The Raman peak marked by ``{*}'' is substrate related.\label{fig:SnOAl2O3}}
\end{figure*}
Three samples, A500, A400, and A300, were grown for $\approx30$\,min
on individual Al$_{\mathrm{2}}$O$\mathrm{_{3}}$(00.1) substrates
at fixed $f_{G}=0.15$~sccm and $T_{g}=$500~$^{\circ}$C, 400~$^{\circ}$C, and 300~$^{\circ}$C,
respectively. Fig.\,\ref{fig:SnOAl2O3} shows structural informaton
of the resulting films obtained by XRD, SEM, and Raman scattering.

\noindent For sample A500, XRD {[}Fig.\,\ref{fig:SnOAl2O3}(a){]}
indicates only the presence of trigonal Sn$_{3}$O$_{4}$,\citep{LAWSON1967}
and possibly SnO$_{2}$ and Sn (whose $2\varTheta$ angles overlap
with those from Sn$_{3}$O$_{4}$). In agreement, bulk sensitive Raman
scattering {[}Fig.\,\ref{fig:SnOAl2O3}(c){]} indicates a strong
contribution from Sn$_{3}$O$_{4}$ but also weak SnO-related peaks.
Surface sensitive Raman scattering {[}Fig.\,\ref{fig:SnOAl2O3}(d){]}
shows additional minor fractions of Sn. The absence of a major fraction
of SnO shown by all these results is in qualitative agreement with
a disproportionation of SnO at temperatures above 410$^{\circ}$C predicted
by the phase diagram (Fig.\,\ref{fig:phasediagram}). Different from
the phase diagram, however, Sn$_{3}$O$_{4}$ instead of SnO$_{2}$
is formed as major oxide phase. The SEM image shows a $\approx300\thinspace$nm-thick,
porous, polycrystalline layer.

Better ordered, $\approx130$\,nm-thick films (A400 and A300) were
obtained at growth temperatures of 400 and 300$^{\circ}$C as shown by the corresponding
SEM images {[}Fig.\,\ref{fig:SnOAl2O3}(b){]}. XRD of these films
{[}Fig.\,\ref{fig:SnOAl2O3}(a){]} shows only the SnO(002) reflex
besides the (00.6) one of the Al$_{2}$O$_{3}$ substrate, indicating
phase-pure SnO, in agreement with the stability region for SnO at
these temperatures in the phase diagram (cf.\,Fig.\,\ref{fig:phasediagram}).
The phase purity is corroborated by the bulk sensitive Raman spectra
{[}Fig.\,\ref{fig:SnOAl2O3}(c){]}. A slightly higher crystal quality
at 400~$^{\circ}$C is signified by the sharper SnO(002) peak and the lower
full width at half maximum (FWHM) of the $\omega$-rocking curve of
the SnO(002) reflex (1.1$^{\circ}$ and 1.9$^{\circ}$ for A400 and A300, respectively,
not shown). The mismatch of rotational symmetry {[}6-fold for the
Al$_{2}$O$_{3}$(00.1) surface and 4-fold for the SnO(001) film{]}
is predicted to result in three rotational domains,\citep{Grundmann2010}
which is indeed reflected by the 12 SnO\{101\} peaks in the skew symmetric
$\Phi$-scan of A400 shown in the inset of Fig.\,\ref{fig:SnOAl2O3}(a).
The three Al$_{2}$O$_{3}$\{10.2\}-peaks in the $\Phi$-scan (indicating
the three-fold bulk rotational symmetry of the corundum structure)
appear at values of $\Phi$ that coincide with \{101\} peaks of the
SnO film on top of the substrate. From these data we can establish
the out-of-plane and in-plane epitaxial relation of SnO on c-plane
Al$_{2}$O$_{3}$ as:

\begin{eqnarray*}
\text{SnO}(001)\text{\,}||\text{\,}\text{Al}{}_{2}\text{O}{}_{3}(00.1) & \text{ for all domains, and}
\end{eqnarray*}

\begin{eqnarray*}
\text{SnO}(100)\,||\,\text{Al}{}_{2}\text{O}{}_{3}(01.0),(-11.0),\text{and}\thinspace(0-1.0)\\
\text{for domains 1--3, respectively.}
\end{eqnarray*}

Interestingly, the surface sensitive Raman spectra {[}Fig.\,\ref{fig:SnOAl2O3}(d){]}
show Sn$_{3}$O$_{4}$ peaks with similar strengths as the SnO-related
ones in A400 and A300, indicating an oxidized surface, likely related
to the cooldown of the film to 200$^{\circ}$C in oxygen plasma after growth.
The stronger Sn$_{3}$O$_{4}$ peak intensity of A400 compared to
that of A300 would also agree with the longer time under oxygen plasma
during cooldown for A400 allowing a deeper oxidation of the SnO surface.
{[}Please note that the relative intensities of the B$_{\text{1g}}$
(115\,cm$^{-1}$) and A$_{\text{1g}}$ (210\,cm$^{-1}$) phonon
lines from SnO are obviously influenced by individual resonance enhancements
occuring for excitation at 2.62 and 3.81\,eV, respectively.{]} The
platelets seen in the SEM image of A400 {[}Fig.\,\ref{fig:SnOAl2O3}(b){]}
show a striking similarity to those of hydrothermally synthesized
Sn$_{3}$O$_{4}$\citep{Manikandan2014}, further corroborating the
assignment of this surface phase.

\subsection{Growth on YSZ(001) at different O-fluxes}

\begin{figure*}
\noindent \centering{}\includegraphics[width=15cm]{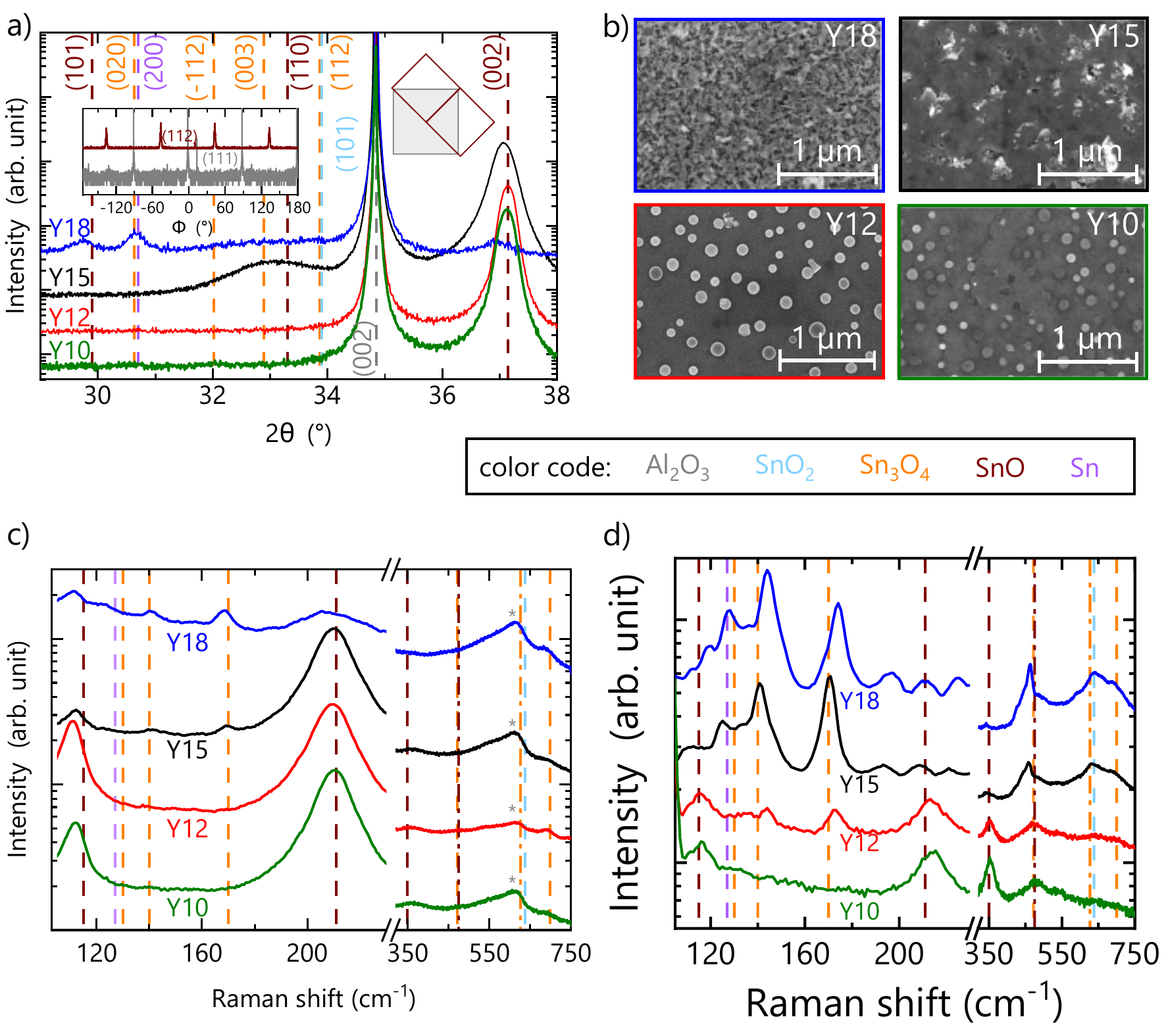}\caption{Structural characterization of SnO films grown on YSZ(001) as function
of oxygen flux $f_{G}$. (a) XRD symmetric out-of-plane $2\varTheta-\omega$
scan. Vertical lines indicate the expected positions of reflexes of
different phases as indicated by color code on the right hand side.
The insets show a $\Phi$-scan of skew symmetric substrate and SnO
reflexes as indicated of the sample grown at 0.12\,sccm, as well
as a pictorial representation of the 45$^{\circ}$ rotated SnO unit cell for
better lattice match. (b) Top view SEM images. (c) and (d) Bulk -and
surface-sensitive Raman spectra measured with an excitation wavelength
of 473 and 325\,nm, respectively. Vertical lines indicate the position
of strong Raman active peaks due to different phases as indicated
by color code on the right hand side. The Raman peak marked by ``{*}''
is substrate related.\label{fig:SnOYSZ}}
\end{figure*}
Based on the highest crystalline quality of the SnO in A400 a growth
temperature of 400~$^{\circ}$C was chosen for the subsequent growth experiments
using YSZ(001) substrates. The four-fold rotational symmetry of this
cubic substrate matches that of the SnO(001) surface and has been
shown to prevent the formation of rotational domains in PLD-grown
films.\citep{SnO_Ogo_Transport,SnO_PLD} Four different oxygen fluxes
(0.18~sccm, 0.15~sccm, 0.12~sccm and 0.10~sccm) in the vicinity
of the stoichiometric flux of 0.15\,sccm (extrapolated in Fig.\,\ref{fig:Oxygen-vs-Growth})
were used to study their impact on phase formation. Fig.\,\ref{fig:SnOYSZ}
shows the structural data of the resulting four, $\approx120$\,nm-thick
films (Y18, Y15, Y12, and Y10, respectively). The sample Y18 (0.18\,sccm)
shows only weak XRD peaks {[}Fig.\,\ref{fig:SnOYSZ}(a){]} likely
related to the polycrystalline, disordered structure visible in the
SEM image {[}Fig.\,\ref{fig:SnOYSZ}(b){]}. The related Raman spectrum
{[}Fig.\,\ref{fig:SnOYSZ}(c){]} shows Sn$_{3}$O$_{4}$-related
peaks and a weak SnO contribution, in agreement with the O-rich flux
stoichiometry. In contrast, the layers grown at lower oxygen flux
(0.15\,--\,0.10 sccm, Y15\,--\,Y10) are showing strong SnO(002)
XRD peaks {[}Fig.\,\ref{fig:SnOYSZ}(a){]} as well as dominant SnO-related
bulk sensitive Raman peaks {[}Fig.\,\ref{fig:SnOYSZ}(c){]}, indicative
of a dominant SnO phase, forming a smooth, compact layer as seen in
the SEM images {[}Fig.\,\ref{fig:SnOYSZ}(c){]}. At an oxygen flux
of 0.15~sccm, however, a broad XRD peak around $2\varTheta=33$\,$^{\circ}$
exists that could be assigned to a secondary Sn$_{3}$O$_{4}$ phase
according to the weak additional bulk-sensitive Raman peaks and likely
visible as crystallites protruding from the film surface. The films
grown at 0.12 and 0.10~sccm, in contrast, are phase pure gauged by
XRD and bulk-sensitive Raman spectra but exhibit droplets visible
in the SEM images, suggesting metallic Sn as a secondary phase. In
the context of the phase diagram Fig.\,\ref{fig:phasediagram} an
O-flux of 0.15\,sccm corresponds to an O-rich stoichiometry with
associated phase ``SnO(s)\,+\,SnO$_{2}$(s)'', whereas 0.12\,sccm
correspond to a Sn-rich stoichiometry with associated phase ``SnO(s)\,+\,Sn(l)''.
Also the films grown on YSZ show that, different from the phase-diagram,
under slightly O-rich conditions the intermediate Sn$_{3}$O$_{4}$
rather than SnO$_{2}$ is formed as major secondary oxide phase.

The SnO(002) XRD reflex of Y15, Y12, and Y10 shows a $\omega$-rocking
curve FWHM of 0.67$^{\circ}$, 0.46$^{\circ}$, and 0.51$^{\circ}$, respectively. The insets of
{[}Fig.\,\ref{fig:SnOYSZ}(a){]} show the $\mathrm{\mathrm{\varPhi}}$\nobreakdash-scan
of Y12 and the related in-plane epitaxial relation between SnO(001)
and YSZ(001) characterized by a 45$^{\circ}$ rotation with respect to each
other, which reduces the mismatch from -34~\% ($a$$_{\mathrm{SnO}}=0.38$~nm,
$a$$_{\mathrm{YSZ}}=0.51$~nm) to 5~\% (2$a$$_{\mathrm{SnO}}=0.76$~nm,
$\sqrt{2}a$$_{\mathrm{YSZ}}=0.72$~nm). These data agree with the
out-of-plane and in-plane epitaxial relation SnO(001)$\,||\text{\,}$YSZ(001)
and SnO(110)$\,||\text{\,}$YSZ(100), respectively, reported in Refs.\,\citenum{SnO_Ogo_Transport,SnO_PLD}
for PLD-grown films.

The surface sensitive Raman spectra shown in {[}Fig.\,\ref{fig:SnOYSZ}(d){]}
indicate pure Sn$_{\mathrm{3}}$O$_{\mathrm{4}}$ for samples Y18
and Y15, dominant SnO with a weak Sn$_{\mathrm{3}}$O$_{\mathrm{4}}$
contribution for Y12, and pure SnO for Y10.

\subsubsection*{Bulk- and surface sensitive Photoelectron Spectroscopy}

\noindent 
\begin{figure*}
\noindent \begin{centering}
\includegraphics[width=7.5cm]{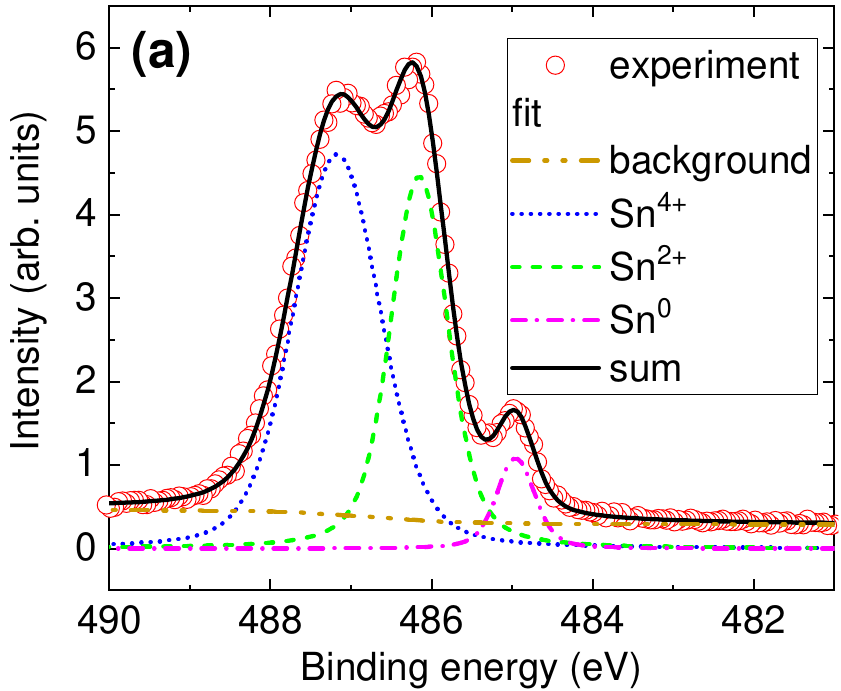}\hspace{1cm}\includegraphics[width=7.5cm]{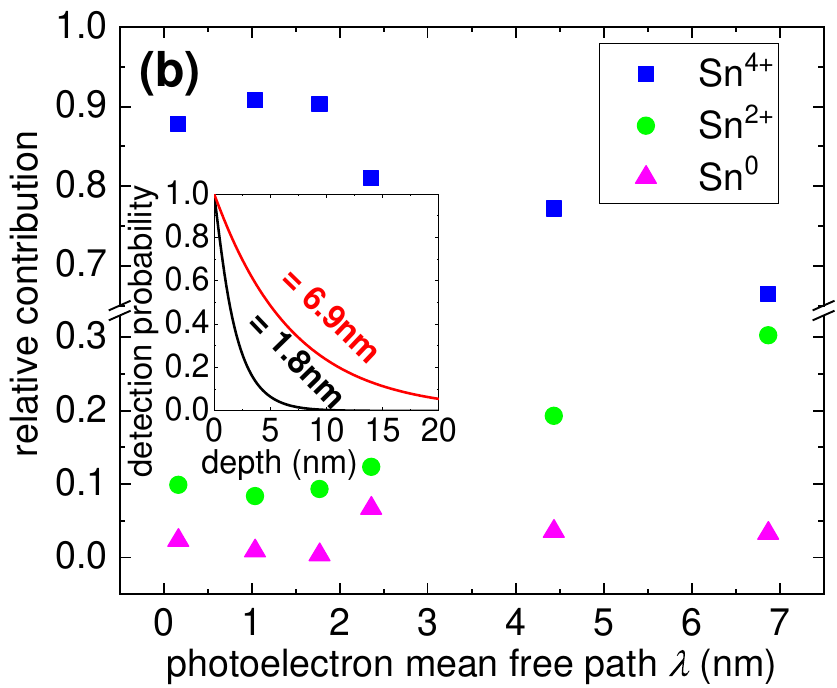}\vspace{0.5cm}
\par\end{centering}
\noindent \begin{centering}
\includegraphics[width=7.5cm]{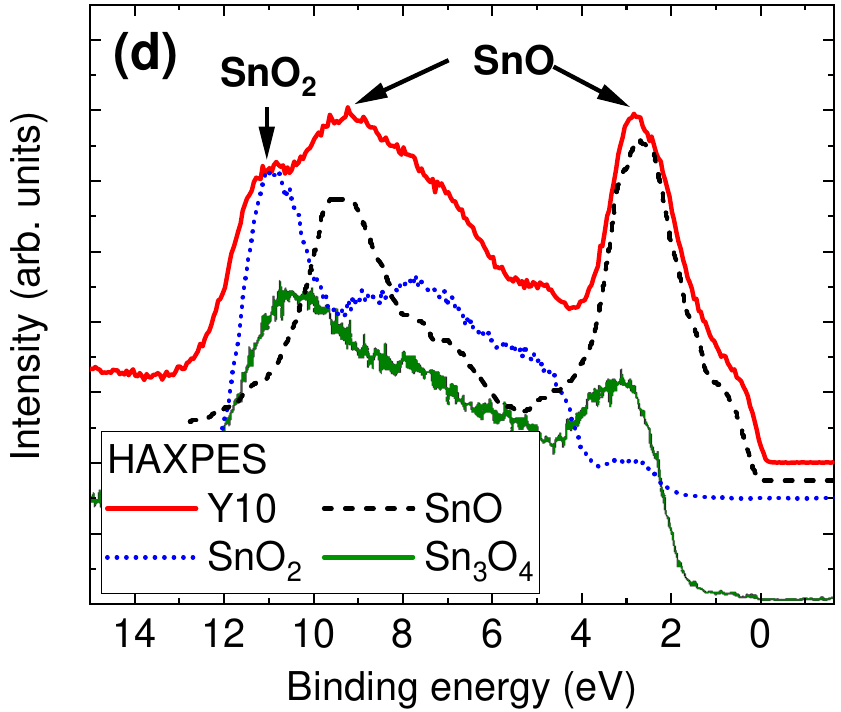}\hspace{1cm}\includegraphics[width=7.5cm]{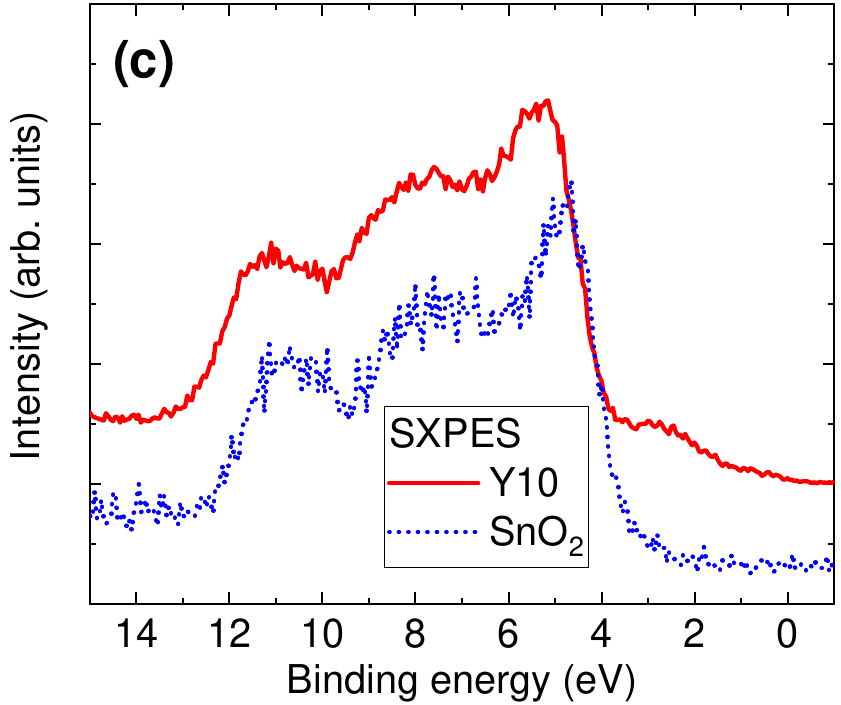}
\par\end{centering}
\noindent \centering{}\caption{Photoelectron spectra of sample Y10. The binding energy scale is calibrated
such that 0\,eV correspond to the Fermi level. (a) Representative
Sn\,3d$_{5/2}$ core level measured by bulk-sensitive HAXPES and
its decomposition into contributions of different Sn valence states
as labeled. (b) Relative contribution of the different Sn-valence
states to the total Sn\,3d$_{5/2}$ core level as function of detection
depth (photoelectron mean free path $\lambda$) measured by SXPES
(0--2\,nm) and HAXPES (2--7\,nm) at different TOA. The inset shows
probability distributions of photoelectron detection as function of
depth for SXPES ($\lambda=1.8$\,nm) and HAXPES ($\lambda=6.9$\,nm)
for the lowest TOAs used in this study (9.7 and 5$^{\circ}$ for SXPES and HAXPES,
respectively).(c) and (d) Valence band spectrum of Y10 (red, solid
line) along with that of a bulk SnO$_{2}$ reference sample (blue
dashed line, taken from Ref.\,\citenum{Nagata2019a}), SnO(001) thin
film grown by PLD taken and adapted from Ref.\,\citenum{SnO_Ogo},
as well as Sn$_{3}$O$_{4}$ taken and adapted from the supporting
information of Ref.\,\citenum{Manikandan2014}. Measurements accomplished
with HAXPES at TOA=5$^{\circ}$ are bulk sensitive (c) while those measured
by SXPES at TOA=9.7$^{\circ}$ are surface sensitive (d).\label{fig:PES}}
\end{figure*}
Sample Y10 was further analyzed by photoelectron spectroscopy. The
Sn\,3d$_{5/2}$ core level, exemplarily shown in Fig.\,\ref{fig:PES}(a),
consists of up to three contributions that are related to an increasing
oxidation state at increasing binding energy. The lowest binding energy
contribution (at $\approx485$\,eV) can be attributed to metallic
Sn (Sn$^{0}$), that at $\approx486$\,\,eV to SnO (Sn$^{2+}$),
and that at $\approx487$\,\,eV to SnO$_{2}$ (Sn$^{4+}$).\citep{Koever1995,Moulder1995,Xia2014}
The intensity of the contributions of the different oxidation states
normalized to the total peak intensity is shown in Fig.\,\ref{fig:PES}(b)
as a function of photoelectron mean free path $\lambda$ ($\lambda<2$\,nm
mesaured by SXPES and $\lambda>2$\,nm by HAXPES). The inset shows
the depth-dependent detection probability for photoelectrons for two
different $\lambda$ that correspond to normal emission HAXPES (6.9\,nm)
and SXPES (1.8\,nm). Even though the detection probability is highest
for photoelectrons from the surface in both cases, HAXPES provides
a larger fraction of photoelectrons from deeper regions than SXPES.
Consequently, Fig.\,\ref{fig:PES}(b) can be discussed as a qualitative
depths profiling of the different Sn-related phases present in the
near-surface region of the film: The dominant contribution is stemming
from Sn\,$^{4+}$ and makes up $\approx90\thinspace\%$ of the peak
intensity within the first 2\,nm, which can only be explained by
a SnO$_{2}$ surface layer. The valence band spectrum taken by surface-sensitive
SXPES resembles that of a pure SnO$_{2}$ bulk sample (taken from
Ref.\,\citep{Nagata2019a}) as shown in Fig.\,\ref{fig:PES}(d).
It thus corroborates the assignment of a few-nm thick SnO$_{2}$ surface
layer, similar to findings on air-exposed SnO MBE-grown from the SnO
vapor.\citep{MBE_SnO_Schlom} With increasing $\lambda$ the Sn\,$^{4+}$
contribution decreases whereas the Sn\,$^{2+}$ contribution steadily
increases from $10\thinspace\%$ at the surface reaching $30\thinspace\%$
at the maximum probed depth ($\lambda=6.9$\,nm). This could in principle
be related to the presence of the mixed-valence tin oxide Sn$_{3}$O$_{4}$
{[}(Sn$^{2+}$)$_{2}$(Sn$^{4+}$)O$_{4}${]}\citep{Manikandan2014}
below the surface. In agreement with the absence of Sn$_{3}$O$_{4}$
in the surface-sensitive Raman spectrum of Y10 {[}Fig.\,\ref{fig:SnOYSZ}(d){]}
the valence band spectrum of Y10 shown in Fig.\,\ref{fig:PES}(c)
taken by bulk-sensitive HAXPES, however, does not match that of the
Sn$_{3}$O$_{4}$ (shown in the supporting information of Ref.\,\citenum{Manikandan2014}).
Instead, the valence band spectrum resembles that of SnO(001) with
a small contribution of SnO$_{2}$ from the surface photoelectrons
as comparison to the reference HAXPES spectra of SnO$_{2}$\citep{Nagata2019a}
and SnO(001)\citep{SnO_Ogo} {[}also shown in Fig.\,\ref{fig:PES}(c){]}
yields. Both contributions are labeled. The shoulder at 1\,eV,
which is also present in HAXPES from the PLD-grown SnO(001)\citep{SnO_Ogo}
may indicate the presence of metallic Sn,\citep{Koever1995} leading
to a valence spectrum similar to that shown for Sn-added $p$-type
SnO films in Ref.\,\citenum{Mohamed2019}. This assignment would
agree with a small but distinct Sn\,$^{0}$ contribution arising
at $\lambda>2$\,nm as shown in Fig.\,\ref{fig:PES}(b). This Sn$^{0}$
contribution makes up $4\thinspace\%$ of the total peak area and
confirms the assigment of Sn metal droplets in the SEM image of Fig.\,\ref{fig:SnOYSZ}(b)
that cover a fraction of the film surface.

Hence, photoelectron spectroscopy of Y10 indicates - different from
surface-sensitive Raman scattering - the presence of a few nm-thick,
fully oxidized (i.e., SnO$_{2}$) layer as well as the dominant presence
of SnO in the bulk, which, however, is mixed with a small fraction
of metallic Sn. Please note the sensitivity of Raman spectroscopy
to SnO$_{2}$ is comparably low with the excitation wavelengths used
in this study. Shorter wavelengths would be required to achieve a
resonant enhancement of the SnO$_{2}$ signal.

\subsection{Summary on phase identification and refined growth window}

\begin{table*}
\caption{Summary of phase identification by the different methods. Phases in
``()'' denote a weak contribution. \label{tab:phases}}
\begin{tabular}{cccccc}
\hline 
method & XRD & Raman & Raman & SEM & PES\tabularnewline
sample &  & bulk & surface &  & \tabularnewline
\hline 
\hline 
A500 & Sn$_{3}$O$_{4}$, (Sn or SnO$_{2}$) & Sn$_{3}$O$_{4}$, (SnO) & Sn$_{3}$O$_{4}$, (SnO, Sn) & porous, polycrystalline & -\tabularnewline
A400 & SnO & SnO, (Sn) & Sn$_{3}$O$_{4}$, SnO, (Sn) & crystallites (+film) & -\tabularnewline
A300 & SnO & SnO & (Sn$_{3}$O$_{4}$, SnO) & film (+ crystallites) & -\tabularnewline
Y18 & (SnO, Sn$_{3}$O$_{4}$ or Sn) & Sn$_{3}$O$_{4}$, (SnO) & Sn$_{3}$O$_{4}$ & polycrystalline & -\tabularnewline
Y15 & SnO, (Sn$_{3}$O$_{4}$) & SnO, (Sn$_{3}$O$_{4}$) & Sn$_{3}$O$_{4}$ & film (+ crystallites) & -\tabularnewline
Y12 & SnO & SnO & SnO, (Sn$_{3}$O$_{4}$) & film (+ Sn droplets) & -\tabularnewline
Y10 & SnO & SnO & SnO & film (+ Sn droplets) & surf. SnO$_{2}$/ bulk SnO(, Sn)\tabularnewline
\hline 
\end{tabular}
\end{table*}
Table\,\ref{tab:phases} gives a summary of the identified phases
in our films by the respective methods. While epitaxial, well-oriented
SnO$_{x}$ phases can be detected by XRD, Raman scattering is an indispensible
tool to also detect polycrystalline SnO$_{x}$ phases with less-well
oriented crystal planes. The major SnO$_{x}$ phase with $x>1$ is
Sn$_{3}$O$_{4}$ rather than SnO$_{2}$ as shown by Raman scattering.
Few-nm thin surface SnO$_{2}$ was only identified by photoelectron
spectroscopy and scanning electron micrographs were decisive in identifying
metallic Sn (droplets). Based on the results we can delineate a growth
window for phase-pure SnO at growth temperatures ranging from 400\,--\,300\,$^{\circ}$C.
Whether it is possible to obtain phase-pure SnO (without additional
Sn) by plasma-assisted MBE remains to be seen. The corresponding growth
window would be delineated by O-fluxes between 0.12\,sccm and 0.15\,sccm
at a growth temperature of 400$^{\circ}$C. Hence, samples Y12 and Y15 are closest
to stoichiometric, single phase SnO samples --- being slightly Sn-rich
and O-rich, respectively.

\section{Electrical, Electrothermal Transport and hole effective mass}

\subsection{Room temperature transport properties}

\noindent 
\begin{table}[h]
\caption{Summary of the Hall measurement results (resistivity $\rho$, Hall
hole concentration\textit{ $p_{H}$,} and Hall mobility $\mu_{H}$)
together with the FWHM ($\triangle\omega$) of the SnO(200) reflex
for all samples. Sample names in parenthesis denote non-SnO films.
Hall measurements failed for A500 and Y18, likely due to a too low
mobility.\label{tab:Hall}}
\begin{tabular}{>{\centering}p{1.7cm}>{\centering}p{1cm}>{\centering}p{1.7cm}>{\centering}p{1.7cm}>{\centering}p{1.7cm}}
\toprule 
sample/

piece & $\mathbf{\boldsymbol{\mathrm{\triangle\omega}}}$\\
($^{\circ}$) & $\rho$\\
 $\mathrm{(\boldsymbol{\Omega cm})}$ & \textbf{\textit{$p_{\text{H}}$}}\\
\textbf{$(\mathrm{10^{18}}$}cm\textbf{$^{-3})$} & \textbf{$\mu_{\text{H}}$}\\
(cm$^{2}$/Vs)\tabularnewline
\midrule
\midrule 
(A500/a) & - & 53 & - & -\tabularnewline
A400/a & 1.10 & 1.32 & $4.8\mathrm{\pm1.0}$ & $1.0\mathrm{\pm0.2}$\tabularnewline
A300/a & 1.87 & 2.08 & $1.8\pm0.3$ & $1.7\pm0.3$\tabularnewline
A300/b &  & 1.10 & $2.3\pm0.1$ & $2.4\pm0.1$\tabularnewline
(Y18/a) & - & 217 & - & -\tabularnewline
Y15/a & 0.67 & 0.9 & $2.5\pm0.1$ & $2.7\pm0.1$\tabularnewline
Y15/b &  & 0.46 & $2.3\pm0.1$ & $6.0\pm0.2$\tabularnewline
Y12/a & 0.46 & 0.4 & $4.1\pm0.1$ & $3.6\pm0.1$\tabularnewline
Y12/b &  & 0.66 & $3.8\pm0.6$ & $2.5\pm0.4$\tabularnewline
Y10/a & 0.51 & 0.25 & $9.7\pm0.8$ & $2.6\pm0.2$\tabularnewline
Y10/b &  & 0.34 & $3.42\pm0.02$ & $5.41\pm0.03$\tabularnewline
\bottomrule
\end{tabular}
\end{table}
The charge carrier transport properties of all films were determined
by room-temperature Hall measurements. Note, that the hole concentration
$p$ and drift mobility $\mu$ are generally related to the quantities
extracted by Hall measurements by the -- often ignored -- Hall (scattering)
factor $r_{H}$ as: $\mu=\mu_{H}/r_{H}$ and $p=p_{H}\cdot r_{H}$.\citep{Sze2007}
The factor $r_{H}$ depends on the charge carrier scattering mechanism,
can range from 1 to 2 in the case of non-degenerate doping, and approaches
unity for degenerate doping. In the non-degenerate case $r_{H}=1.93$
for ionized impurity scattering,\citep{Sze2007} and has been calculated
to be $r_{H}=1.77$ for phonon-limted transport in SnO.\citep{Hu2019}
For better comparison to literature results we are initially discussing
the Hall quantities $p_{H}$ and $\mu_{H}$ of our samples shown in
Table\,\ref{tab:Hall}. All films with dominant SnO(001) identified
by XRD showed $p$-type conductivity with varying Hall hole concentration
($p_{H}$ between $1.8\times10^{18}$ and $9.7\times10^{18}$\,cm$^{-3}$)
and resistivities $\rho$ on the order of 1\,$\Omega$cm, whereas
the non-SnO films were highly resistive ($\rho>50\thinspace\Omega$cm).
A spread of transport properties on different pieces (named (``a'',
``b'', ...) from the same grown wafer is likely related to slight
flux and/or temperature inhomogeneities, highlighting the optimization
potential by finetuning growth parameters similar to what has been
reported in Refs.\,\citenum{Minohara2019,Minohara2019a,Record_mob_SnO}.

\noindent The lower hole mobility in the films on sapphire compared
to those on YSZ is likely related to the rotational domains and associated
domain-boundary scattering. For the single-crystalline films on YSZ,
the highest Hall mobility of $\mu_{H}=6.0$\,cm$^{2}$/Vs is found
in Y15. 

\subsection{Temperature-dependent transport, scattering mechansim, and acceptor
type}

\noindent We investigated the two samples Y15 and Y12 that are closest
to stoichiometric, phase-pure SnO, additionally by temperature-dependent
Hall measurements between 350 and 100\,K, and discuss the result
in the context of available literature data. Analyzing the temperature
dependence of transport properties allows us to conclude on transport
mechanisms, acceptor ionization energy, and scattering mechanism.
To date, the limited number of reports on temperature-dependent thin-film
transistor characteristics\citep{Fortunato_SnO,Zhang_SnO_Transport,Kim_SnO-Transport}
and temperature-dependent Hall measurements \citep{SnO_Ogo_Transport,SnO_PLD,Minohara2019a,Becker2019}
of unintentionally-doped SnO show a variety of different transport
characteristics: The observation of decreasing hole mobility with
decreasing temperature has been typically associated with hopping
conductivity or percolative transport,\citep{SnO_Ogo_Transport,Kim_SnO-Transport,Fortunato_SnO,Zhang_SnO_Transport}
whereas an increasing hole mobility has been associated with phonon-scattering
limited band transport by free holes.\citep{Kim_SnO-Transport,Miller2017}

\noindent 
\begin{figure}
\noindent \begin{centering}
\includegraphics[width=8.5cm]{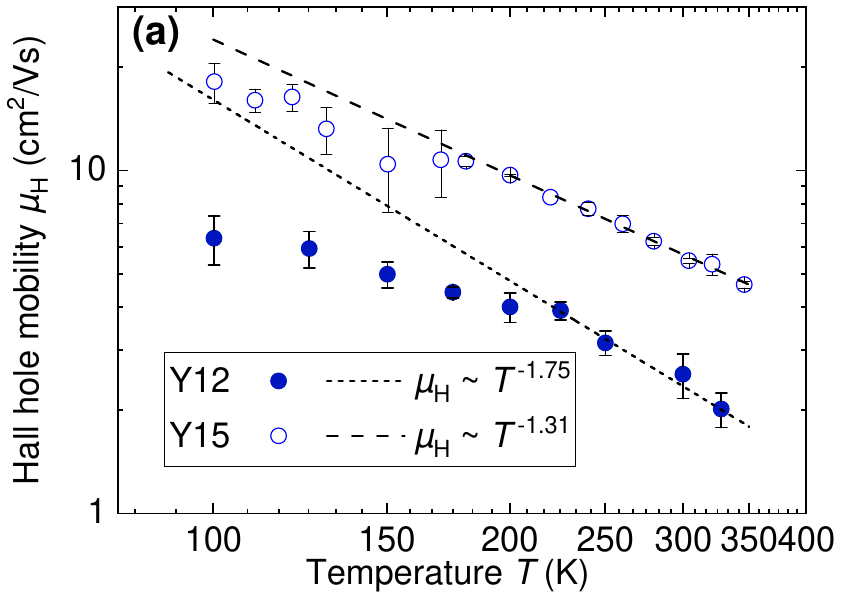}
\par\end{centering}
\noindent \begin{centering}
\includegraphics[width=8.5cm]{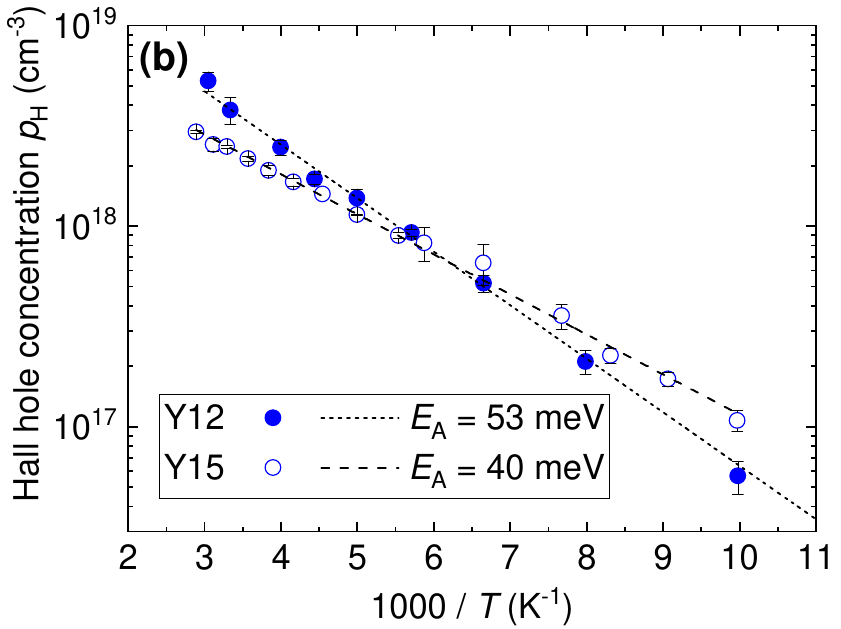}
\par\end{centering}
\caption{Hole transport properties of Y12/b and Y15/b determined by van-der-Pauw-Hall
measurements at temperatures $T$ in the range of 100 to 330~K. (a)
Hall hole mobility $\mu_{H}$ in a double-logarithmic representation
including a fitting curve of the high-temperature region. (b) Hall
hole concentration $p_{H}$ in an Arrhenius-type representation with
associated fit and apparent activation energy $E_{a}$.\label{fig:TdepHall}}
\end{figure}
 Fig.\,\ref{fig:TdepHall}(a) shows the extracted Hall hole mobility
in Y12/b and Y15/b as function of temperature in a double-logarithmic
representation. From room temperature to 100\,K the mobilities of
Y12/b and Y15/b increase from 2.5\,and\,6.0\,cm$^{2}$/Vs to 6\,and\,18\,cm$^{2}$/Vs,
respectively. The strongly increasing mobility with decreasing temperature
clearly indicates dominant phonon scattering, in both samples, thus
excluding hopping transport as well as a dominant influence of ionized
impurity scattering (e.g. due to high compensation) or other mobility-limiting
scattering mechanisms (dislocation, neutral impurity) at room temperature.
In fact, the observed room-temperature mobilities of Y12/b and Y15/b
are above the estimated mobility limit of $\mu_{\text{H}}^{\text{IIS}}\approx2.3$\,cm$^{2}$/Vs
for ionized impurity scattering, taking into account $r_{H}=1.93$
for IIS and the drift mobility $\mu^{\text{IIS}}\approx1.2$\,cm$^{2}$/Vs
predicted for an ionized impurity concentration equalling our hole
concentration in Y12/b,\citep{Minohara2019} further ruling out significant
ionized impurity scattering and compensation. Similar to Ref.\,\citenum{Kim_SnO-Transport}
we assign optical phonon scattering to the behavior above 200\,K
due to the large exponent of the observed $\mu_{\text{H}}\propto T^{-1.31}$
and $\mu_{\text{H}}\propto T^{-1.75}$-dependence of Y15/b and Y12/b,
respectively. In addition, first principles calculations predict
phonon-scattering limited Hall hole mobilities of $\mu_{\text{H}}^{\text{POP}}=106$
and $13$\,cm$^{2}$/Vs in the $c$-direction and perpendicular
to this direction (i.e., in the (001) plane), respectively,\citep{Hu2019}
being mainly limited by polar optical phonon scattering rather than
by acoustic phonon scattering. Its anisotropy is mainly related to
the anisotropy of the effective hole mass predicted by first principles
calculations to be $m_{[001]}^{*}\approx0.55\mathrm{m_{e}}$ along
the {[}001{]} direction and around $m_{[100]}^{*}=m_{[010]}^{*}\approx3.0\mathrm{m_{e}}$
in the (001)-plane with free electron mass $m_{e}$.\citep{Varley2013,Ha2017}
The phonon-limited, room-temperature hole mobilities at of Y12/b and
Y15/b are ranging between reported experimental Hall hole mobilities
from 2 to 3\,cm$^{2}$/Vs (at Hall hole concentrations in the range
of $2.5\cdot10^{16}$ to $2.5\cdot10^{17}$\,cm$^{-3}$) on other
(001)-oriented, single crystalline layers,\citep{MBE_SnO_Schlom,SnO_Ogo_Transport,SnO_PLD}
and the theoretically predicted limit in the (001)-plane. On the other
hand, significantly higher room-temperature Hall mobilities have been
reported for polycrystalline SnO ($\mu_{\text{H}}=30$~cm$^{2}$/Vs\citep{Miller2017}
at $p_{H}=7\cdot10^{15}$\,cm$^{-3}$, $\mu_{\text{H}}=19$~cm$^{2}$/Vs\citep{Record_mob_SnO}
at $p_{\text{H}}=2.2\cdot10^{17}$\,cm$^{-3}$), which can be understood
in terms of transport through crystallites with the $c$-axis (high-mobility
direction) oriented in-plane. Interestingly, Minohara et al. reported
high hole mobilities for single-crystalline (001)-oriented SnO layers
(c-axis out-of-plane) ($\text{\ensuremath{\mu_{\text{H}}=15}}$~cm$^{2}$/Vs\citep{Minohara2019a}
at $p_{\text{H}}=10^{17}$\,cm$^{-3}$ and $\text{\ensuremath{\mu_{\text{H}}=21}}$~cm$^{2}$/Vs\citep{Minohara2019}
at $p_{\text{H}}=7\cdot10^{16}$\,cm$^{-3}$), which is difficult
to explain. In addition, the decreasing Hall hole mobility with decreasing
temperature reported for these layers\citep{Minohara2019a} indicates
that the room-temperature mobility is mainly limited by other scattering
mechanisms than phonon scattering. Since phonon scattering cannot
be avoided at room temperature, the high mobility may be related to
a different transport mechanism, e.g. with significantly lower effective
hole mass.

\noindent Fig.\,\ref{fig:TdepHall}(b) shows the extracted Hall hole
concentration of samples Y12/b and Y15/b as function of temperature
in an Arrhenius type plot. It decreases with decreasing temperature
following an activated behavior at an activation energy of $E_{A}=53$\,and\,40\,meV,
respectively. In fact, most works on unintentionally-doped SnO report
an activated Hall hole concentration with room-temperature values
ranging from $7\cdot10^{15}$ to $2.5\cdot10^{17}$\,cm$^{-3}$ and
apparent activation energies $E_{A}$ in the range from 220 to 40\,meV,\citep{Miller2017,SnO_Ogo_Transport,SnO_PLD}
indicative of a non-degenerate doping concentration. Assuming a density-of-state
(DOS) effective hole mass of $m_{h}^{*}=(m_{[100]}^{*}\cdot m_{[010]}^{*}\cdot m_{[001]}^{*})^{1/3}\approx1.7m_{e}$
(using the anisotropic effective mass from Ref.\,\citep{Varley2013}),
a relative permitivity of $\epsilon_{r}=18.8$,\citep{Li2015} and
the hydrogenic Bohr radius $a_{B}=0.053$\,nm the Mott criterion\citep{Mott1982}
predicts a critical hole density

\noindent 
\begin{eqnarray}
p_{\text{Mott}} & = & [(0.26\cdot m_{h}^{*})/(\epsilon_{r}\cdot a_{B}\cdot m_{e})]^{3}\label{eq:Mott}
\end{eqnarray}

\noindent of $p_{\text{Mott}}\approx9\cdot10^{19}\text{\,\text{cm}}{}^{-3}$,
which indicates a non-degenerate doping concentration for all SnO
layers discussed in this work and the cited literature. Notwithstanding,
Minohara et al. report a temperature-independent Hall hole concentration
of $p_{H}\approx10^{17}$\,cm$^{-3}$ {[}from which we estimated
$E_{A}\approx(-1\pm2)$\,meV{]},\citep{Minohara2019a} indicative
of a (potentially highly compensated) degenerate acceptor concentration,
suggesting distinctly different conduction mechanism from that in
Y12/b, Y15/b, and most other reported literature.

\begin{figure}
\noindent \begin{centering}
\includegraphics[width=8.5cm]{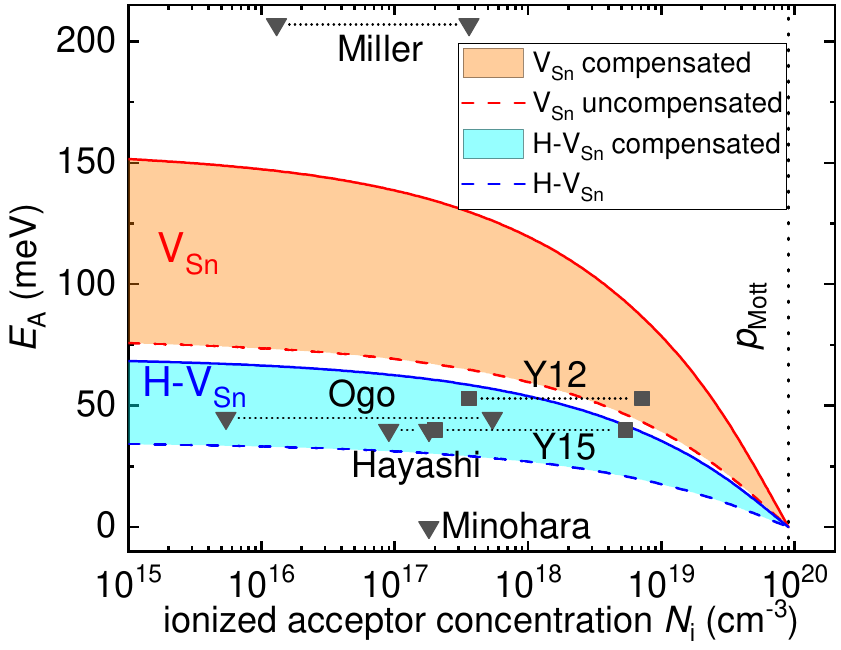}
\par\end{centering}
\caption{Comparison of the experimentally obtained apparent activation energy
$E_{A}$ (symbols) to theoretically prediced ones for V$_{\text{Sn}}$
and H-V$_{\text{Sn}}$\citep{Varley2013} as a function of ionized
acceptor concentration $N_{i}$ taking into account the impact of
impurity bands and compensation. Experimental data points are plotted
under the assumption of no compensation ($N_{i}=p$) and would need
to be shifted to higher $N_{i}$ in case of compensation. Symbol labels
indicate the source of data: ``Miller'',\citep{Miller2017} ``Ogo'',\citep{SnO_Ogo_Transport}
``Hayashi'',\citep{SnO_PLD}, ``Minohara'',\citep{Minohara2019a}
``Y12'' and ``Y15'' samples Y12/b and Y15/b of the present work.\label{fig:EAct}}
\end{figure}
For non-degenerately doped material, the assignment of the acceptor
type is commonly based on the value of the activation energy $E_{A}$.
Using first principles calculations Varley et al. predicted isolated
tin vacancies (V$_{\text{Sn}}$) and their complex with hydrogen (H-V$_{\text{Sn}}$)
to be potential acceptors with (0/-1) charge transition levels 155
and 70~meV above the valence band maximum, respectively, that can
cause free holes in unintentionally-doped SnO.\citep{Varley2013}
These energies are considered to be identical with the acceptor ionization
energy $\epsilon_{A}^{0}$ for isolated acceptors. With increasing
ionized acceptor concentration $N_{i}$, this ionization energy decreases
according to $\epsilon_{A}=\epsilon_{A}^{0}[1-(N_{i}/p_{\text{Mott}})^{1/3}]$\citep{Pearson1949}
due to the onset of acceptor band formation, and disappears at the
critical density $N_{i}=p_{\text{Mott}}$.\citep{Grundmann2006} The
apparent activation energy extracted from the temperature-dependent
hole concentration relates to $\epsilon_{A}$ by $E_{A}=\epsilon_{A}$
in the case of compensated doping or $E_{A}=\epsilon_{A}/2$ in the
case of uncompensated doping.\citep{Grundmann2006} Fig.\,\ref{fig:EAct}
compares the experimentally obtained $E_{A}$ from Y12/b and Y15/b
as well as Refs.\,\citenum{SnO_PLD, Miller2017, Minohara2019a,SnO_Ogo}
to the theoretically predicted ones for V$_{\text{Sn}}$ and H-V$_{\text{Sn}}$
taking into account the effect of acceptor band formation.  (Note,
that we calculated $p$ from measured and reported $p_{H}$ assuming
a Hall factor of $r_{H}=1.8$ to reflect dominant polar optical phonon
scattering.) Given that the mobility behavior suggests rather uncompensated
acceptors, the results for Y12/b and Y15/b match the case of V$_{\text{Sn}}$
better than that of H-V$_{\text{Sn}}$. The large $E_{A}$ extracted
from data of Ref.\,\citenum{Miller2017}, on the other hand cannot
be explained by either V$_{\text{Sn}}$ or H-V$_{\text{Sn}}$.\footnote{We note that the activation energy of $\approx0.09$\,eV stated in
reference \citenum{Miller2017} does not match the data presented
in that paper. We believe that it was likely erroneously extracted
from an Arrhenius-plot using $\log p$ instead of $\ln p$.}

\subsection{Electrothermal transport and DOS effective hole mass}

\begin{table*}
\caption{Calculated DOS effective hole mass (\textbf{$m_{h}^{*}$}) based on
$p_{H}$ and $S$ from Hall and Seebeck measurements for all SnO samples
as data from literature using Eq.\,\ref{eq:Snd} and the related
formalism described in Ref.\,\citenum{Preissler2013}. A Hall factor
of $r_{H}=1.8$ is assumed. The limiting cases polar optical phonon
scattering (POPS, $r=0.5$) and ionized impurity scattering (IIS,
$r=1.5$) are considered. For comparison, the hole concentration $p_{SPH}$
calculated from $S$ using Eq.\,\ref{eq:SSPH} is given to test the
hypothesis of transport by small polaron hopping.\label{tab:SnO_Seebeck_m}}

\noindent \centering{}\renewcommand{\arraystretch}{1.0}%
\begin{tabular}{>{\centering}p{1.8cm}>{\centering}p{2.2cm}>{\centering}p{2.3cm}>{\centering}p{1.4cm}>{\centering}p{2.2cm}>{\centering}p{2.4cm}>{\centering}p{2.4cm}}
\toprule 
sample/

piece & \textit{p$_{H}$}\\
 ($\boldsymbol{\mathrm{10^{18}}\mathrm{/cm^{3}}}$) & \textit{p}\\

($\boldsymbol{\mathrm{10^{18}}\mathrm{/cm^{3}}}$) & {\small{}$\mathbf{S}$} \\
(\textmu V/K) & \textit{p$_{\mathrm{\boldsymbol{SPH}}}$}\\
 ($\boldsymbol{\mathrm{10^{18}}\mathrm{/cm^{3}}}$) & $\boldsymbol{m}_{\boldsymbol{h}}^{\boldsymbol{*}}$ ($\mathrm{\boldsymbol{m}_{\boldsymbol{e}}}$)\\
POPS, $r=0.5$ & $\boldsymbol{m}_{\boldsymbol{h}}^{\boldsymbol{*}}$($\mathrm{\boldsymbol{m}_{\boldsymbol{e}}}$)\\
IIS, $r=1.5$\tabularnewline
\midrule
\midrule 
(A500/a) & - & - & \textcolor{brown}{-266$\pm$7} & - & - & -\tabularnewline
A400/a & 4.8 $\mathrm{\pm}$ 1.0 & 8.6 $\mathrm{\pm}$ 1.80 & 616$\pm$26 & 42 $\mathrm{\pm}$ 14 & 7.94 & 4.10\tabularnewline
A300/a & 1.8 $\mathrm{\pm}$ 0.3 & 3.24 $\mathrm{\pm}$ 0.54 & 597$\pm$17 & 52 $\mathrm{\pm}$ 11 & 3.58 & 1.85\tabularnewline
(Y18/a) & - & - & - & - & - & -\tabularnewline
Y15/a & 2.5 $\mathrm{\pm}$ 0.1 & 4.50 $\mathrm{\pm}$ 0.18 & 480$\pm$45 & 228 $\mathrm{\pm}$ 109 & 1.82 & 0.96\tabularnewline
Y12/b & 3.8 $\mathrm{\pm}$ 0.6 & 6.8 $\mathrm{\pm}$ 1.1 & 523$\pm$7 & 122 $\mathrm{\pm}$ 10 & 3.33 & 1.74\tabularnewline
Y10/a & 9.7 $\mathrm{\pm}$ 0.8 & 17.46 $\mathrm{\pm}$ 1.44 & 543$\pm$9 & 97 $\mathrm{\pm}$ 10 & 7.27 & 3.78\tabularnewline
Hayashi\cite{SnO_PLD} & 0.1 & 0.18 & 763 & 7.56 & 1.88 & 0.96\tabularnewline
Hosono\cite{Hosono2011} & 0.71 & 1.28 & 479 & 202.9 & 0.78 & 0.41\tabularnewline
Miller\citep{Miller2017} & 0.013 & 0.023 & 630 & 35.3 & 0.17 & 0.09\tabularnewline
Becker\cite{Becker2019} & 0.0046 & 0.0083 & 550 & 89.3 & 0.05 & 0.02\tabularnewline
Ogo\citep{SnO_Ogo} & 0.25 & 0.45 & 1990 & $2\times10^{-6}$ & 45766 & 23497\tabularnewline
\bottomrule
\end{tabular}\\
\end{table*}
The room-temperature Seebeck coefficient $S$ measured for all samples
is shown in Tab.\,\ref{tab:SnO_Seebeck_m}. Its positive value further
confirms hole conduction in all SnO samples (A300, A400, Y15, Y12,
Y10), whereas A500 exhibits a negative $S$-- in agreement with the
reported $n$-type conductivity of Sn$_{3}$O$_{4}$.\citep{Suman2014}
The high resistivity of Y18 did not allow for a reliable Seebeck measurement.

The Seebeck coefficient is related to the bulk carrier concentration
and can be used as an alternative to Hall measurements for the estimation
of $p$ if the transport mechanism is known. For example, in oxides
with hole transport by small-polaron hopping $S$ can be related to
$p$ by:

\begin{eqnarray}
S_{SPH} & = & \frac{k_{B}}{e}\ln[(2-2c)/c]\label{eq:SSPH}
\end{eqnarray}
with Boltzmanns constant $k_{B}$, electronic charge $e$, and the
fraction of occupied carrier sites $c=p/N$, i.e., the ratio of hole
concentration $p$ to concentration $N$ of sites that can be occupied
by a hole (e.g., the Sn-site in SnO).\citep{Farrell2015} For example,
this relation has been used to estimate $p$ from the measured $S$
in doped $p$-type oxides whose hole mobility is too low ($\text{\ensuremath{\mu} \ensuremath{\ll1}}$\,cm$^{2}$/Vs)
to allow for Hall measurements, i.e., in Cr$_{2}$O$_{3}$:Mg,\citep{Farrell2015}
LaCrO$_{3}$:Sr,\citep{Zhang2015} and NiO:Li.\citep{Zhang2018} For
the band\nobreakdash-like transport a different relation $S=S(p,m_{h}^{*},r)$
holds which has an additional dependence on the DOS effective hole
mass (\textbf{$m_{h}^{*}$}) and the Seebeck scattering parameter
($r$). For non-degenerate doping (which applies to our films, since
$p\ll p_{\text{Mott}}$) it reads as:\citep{Seeger2004}
\begin{eqnarray}
S_{nd} & = & \frac{k_{B}}{e}\cdot\left(r+\frac{5}{2}-\frac{E_{VBM}-E_{F}}{k_{B}T}\right)\label{eq:Snd}
\end{eqnarray}
with the term ($E_{VBM}-E_{F}$) denoting the distance between Fermi
level $E_{F}$ and valence band maximum $E_{VBM}$ being related to
the hole concentration $p$ through semiconductor statistics (as described
in detail in Ref.\,\citenum{Preissler2013}) using the valence band
DOS parametrized by $m_{h}^{*}$. The Seebeck scattering parameter
varies between $r=-0.5$ for dominant acoustic phonon scattering and
$r=1.5$ for dominant ionized impurity scattering. Optical phonon
scattering is typically described by a scattering parameter of $r=0.5$.\citep{Ginley2011}

Since no experimental values of $m_{h}^{*}$ of SnO have been published
to date we are using the combination of $p_{H}$ determined by Hall
measurements and measured Seebeck coefficient $S$ to estimate $m_{h}^{*}$
based on Eq.\,\ref{eq:Snd} for the different $r$ as previously
demonstrated for the $n$-type semiconducting oxide In$_{2}$O$_{3}$.\citep{Preissler2013}
The results are shown in Tab.\,\ref{tab:SnO_Seebeck_m} along with
the hypothetical hole concentration $p_{SPH}$ derived from $S$ under
the assumption of small polaron hopping using Eq.\,\ref{eq:SSPH}
and $N=2.66\cdot10^{22}$\,cm$^{-3}$, the concentration of Sn-atoms.
For comparison we have added $p_{H}$ and associated $S$ reported
in the literature.\citep{SnO_PLD,Becker2019,Hosono2011,Miller2017,SnO_Ogo}
The drastic discrepancy of the extracted $p_{SPH}$ assuming small
polaron hopping and measured $p_{H}$ demonstrates for all samples,
that the transport is not well described by small polaron hopping.
This corroborates the assumption of band transport by free holes and
consequently, the applicability of the used model according to Eq.\,\ref{eq:Snd}.
Assuming hole transport to be mainly limited by polar optical phonon
scattering ($r=0.5)$ or ionized impurity scattering ($r=1.5)$ we
extracted values of the DOS effective hole mass $m_{h}^{*}$ between
$\approx1$\,$m_{e}$ and $\approx8$\,$m_{e}$ for SnO our films and that of Ref.\,\citenum{SnO_PLD}.
These values are in fair agreement with the theoretically predicted
$m_{h}^{*}=1.7$\,$m_{e}$, and can be seen as an experimental confirmation
of its order of magnitude. Published $p_{\text{H}}$ and $S$ from
Refs.\,\citenum{Hosono2011, Miller2017, Becker2019}, would result
in significantly lower values of $m_{h}^{*}$. This discrepancy may
be related to different transport mechanism or an inhomogeneous carrier
distribution, for example due to the confinement in a thin accumulation
layer. The actual hole concentration in the accumulation layers would
be higher than that extracted from Hall measurements (that assume
the carriers to be spread across the entire film thickness),\citep{Papadogianni2015}
and would consequently lead to a larger extracted DOS effective hole
mass.

\section{Stability of the SnO phase after growth}

The thermal stability of SnO films is highly relevant for their application
with respect to the temperature budget during device processing, e.g.,
contact annealing, and the operation temperature, e.g., in power devices.
An early work by Moh reports SnO to be stable with respect to disproportionation
or oxidation only up to 270$^{\circ}$C.\citep{Moh1974} In contrast, a number
of later publications describe the transformation of polycrystalline
SnO into SnO$_{2}$ upon annealing in different atmospheres to proceed
at temperatures between 400~$^{\circ}$C and 550~$^{\circ}$C.\citep{Oyabu1982,Geurts1984,Das1987,Reddy1989,Yabuta2010,Lin2017}:
Geurts et al. describe the oxidation process from SnO to SnO$_{2}$
through the intermediate stoichiometries e.g., Sn$_{3}$O$_{4}$ or
Sn$_{2}$O$_{3}$ at temperatures ranging from 450 to 650$^{\circ}$C to start
by an internernal displacement of oxygen (disproportionation) followed
by oxidation through incorporation of external oxygen.\citep{Geurts1984}.
For the complete oxidation to SnO$_{2}$ Reddy et al. reported a temperature
of 600~$^{\circ}$C during a two hour annealing in oxygen.\citep{Reddy1989}
On the other hand, Pei et al. reported highly stable SnO layers reaching
their highest crystalline quality during RTA (with unspecified annealing
time) at 700~$^{\circ}$C in nitrogen.\citep{SnO_Ebeam} Interestingly, Yabuta
et al. demonstrated that a SiO$_{x}$ capping layer preseves the SnO
layer by preventing oxygen exchange with the environment using annealing
experiments in nitrogen, oxygen and air at 400~$^{\circ}$C.\citep{Yabuta2010}
We note, however, that a capping layer cannot prevent disproportionation
of the film at temperatures above $\approx400$\,$^{\circ}$C (cf. Fig\ref{fig:phasediagram}).
\begin{figure}
\noindent \begin{centering}
\includegraphics[width=8.5cm]{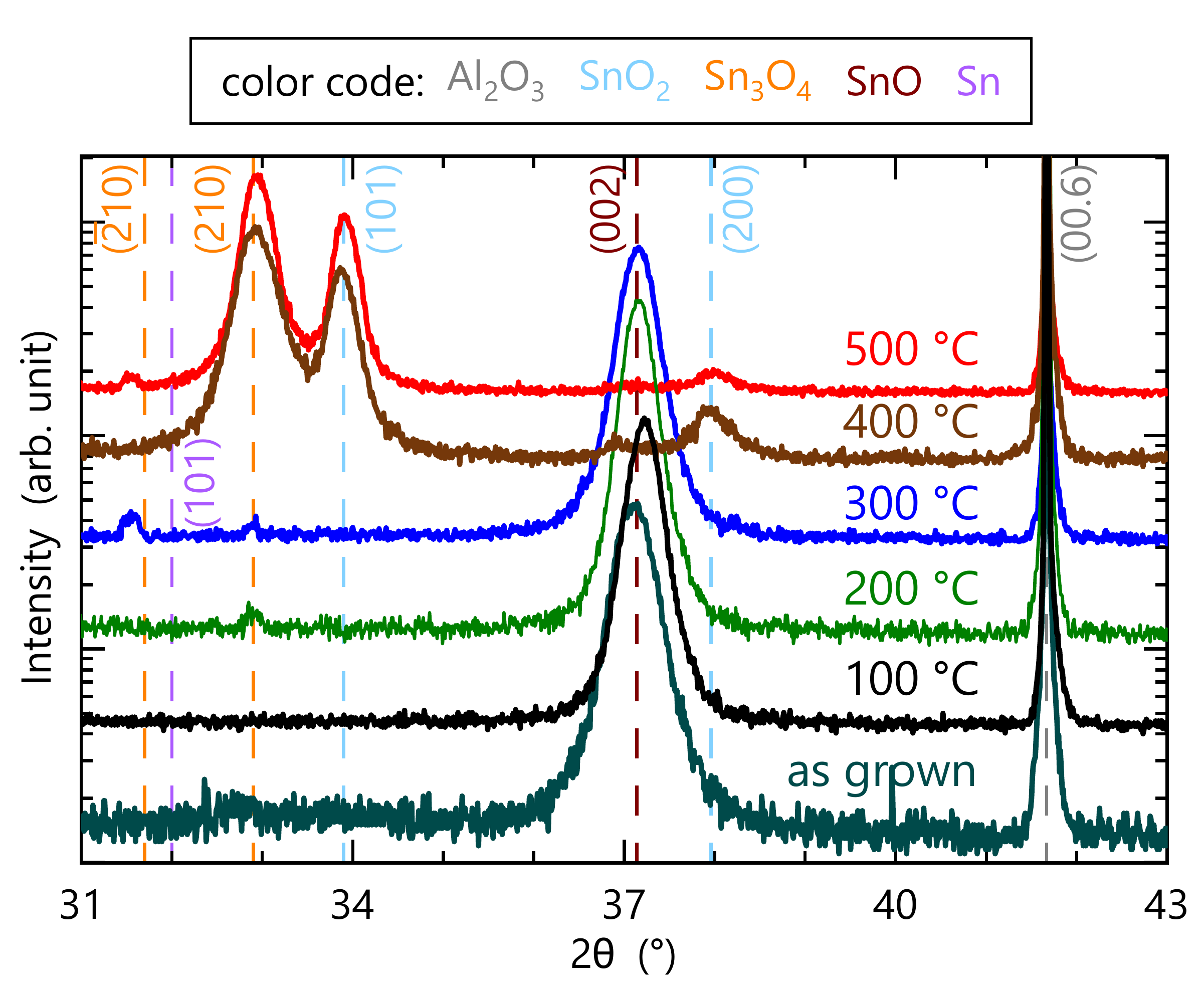}
\par\end{centering}
\caption{Symmetric XRD $2\theta-\omega$ scans of pieces of A400 annealed in
nitrogen at temperatures ranging from 100~$^{\circ}$C to 500~$^{\circ}$C. For reference
purposes the scan of the as grown A400 is shown.\label{fig:Annealing_XRD_SnO}}
\end{figure}

We investigated the thermal stability our SnO films with the example
of samples A400 and A400{*}. Sample A400{*} is an additional sample
grown under the same growth conditions as A400, showing the same XRD
reflexes and exhibiting similar transport properties. Different pieces
of these samples were annealed by RTA at temperatures between 100~$^{\circ}$C
and 500~$^{\circ}$C for 10 minutes in nitrogen, oxygen and forming gas ($\mathrm{N_{2}+H_{2}}$)
atmospheres. In Fig.~\ref{fig:Annealing_XRD_SnO} symmetric out-of-plane
XRD $2\theta-\omega$ scans of A400 annealed in nitrogen at various
temperatures shows preservation of the SnO layer up to 300$^{\circ}$C and its
transformation into mainly Sn$_{3}$O$_{4}$\citep{Balgude2019} and
SnO$_{2}$ for temperatures of 400~$^{\circ}$C and above, in good agreement
with the phase diagram in Fig.\,\ref{fig:phasediagram}. The effect
of annealing treatments in different atmospheres on the transport
properties is shown in Tab.~\ref{tab:Hall_Annealing_SnO_gases}.
Irrespective of the atmosphere, $P$-type conductivity is preserved
for annealing at 300$^{\circ}$C and below with only small quantitative changes
that may also be related to inhomogeneity across the wafer. More significantly,
\textit{n}\nobreakdash-type conductivity is observed after annealing
at 400~$^{\circ}$C and above for all tested atmospheres, in agreement with
the phase change seen by the XRD results of the samples annealed in
N$_{2}$.%
\begin{table}
\caption{Results of Hall measurements on pieces of A400 and A400{*} annealed
in nitrogen (N$_{2}$), oxygen (O$_{2}$) and forming gas (FG) at
different temperatures. Annealing temperature $T_{ann.}$ (``a.g.''
denotes as-grown, the ``{*}'' denotes pieces of sample A400{*}),
resistivity $\rho$, Hall hole concentration $p_{H}$. (negative values
denote $n$-type conductivity), and Hall hole mobility $\mu_{H}$.\label{tab:Hall_Annealing_SnO_gases}}

\noindent \centering{}\renewcommand{\arraystretch}{1.0}%
\begin{tabular}{>{\centering}p{1.5cm}>{\centering}p{1cm}>{\centering}p{1cm}>{\centering}p{2cm}>{\centering}p{2cm}}
\toprule 
$\mathrm{\boldsymbol{T}_{\boldsymbol{ann.}}}$

($^{\circ}$C) & gas & $\boldsymbol{\rho}$ \\
($\boldsymbol{\mathrm{\Omega cm}}$) & \textit{p$_{H}$} \\
($\boldsymbol{10^{18}\mathrm{cm^{-3}}}$) & $\boldsymbol{\mu_{H}}$\\

($\boldsymbol{\mathrm{cm^{2}}\mathrm{/Vs}}$)\tabularnewline
\midrule
\midrule 
a.g. & - & 1.0 & 3.9 $\mathrm{\pm}$ 0.6 & 1.7 $\mathrm{\pm}$ 0.3\tabularnewline
a.g.{*} & - & 1.9 & 2.7 $\mathrm{\pm}$ 0.3 & 1.2 $\mathrm{\pm}$ 0.3\tabularnewline
100 & N$_{2}$ & 0.8 & 6.5 $\mathrm{\pm}$ 0.8 & 1.2 $\mathrm{\pm}$ 0.2\tabularnewline
200 & N$_{2}$ & 0.6 & 6.8 $\mathrm{\pm}$ 1.3 & 1.5 $\mathrm{\pm}$ 0.3\tabularnewline
300{*} & N$_{2}$ & 1.2 & 3.5 $\mathrm{\pm}$ 0.2 & 1.5 $\mathrm{\pm}$ 0.1\tabularnewline
300{*} & O$_{2}$ & 1.3 & $3.6\pm0.5$\textcolor{cyan}{} & $1.3\pm0.2$\textcolor{cyan}{}\tabularnewline
300{*} & FG & 1.1 & 2.5 $\mathrm{\pm}$ 0.1 & 2.2 $\mathrm{\pm}$ 0.1\tabularnewline
\textcolor{brown}{400} & \textcolor{brown}{N$_{2}$} & \textcolor{brown}{0.1} & \textit{\textcolor{brown}{-}}\textcolor{brown}{{} 40 $\mathrm{\pm}$
4} & \textcolor{brown}{1.0 $\mathrm{\pm}$ 0.1}\tabularnewline
\textcolor{brown}{400{*}} & \textcolor{brown}{O$_{2}$} & \textcolor{brown}{0.2} & \textit{\textcolor{brown}{-}}\textcolor{brown}{1.9 $\mathrm{\pm}$
0.2} & \textcolor{brown}{7.3 $\mathrm{\pm}$ 0.7}\tabularnewline
\textcolor{brown}{400{*}} & \textcolor{brown}{FG} & \textcolor{brown}{0.5} & \textit{\textcolor{brown}{-}}\textcolor{brown}{5.8 $\mathrm{\pm}$
0.6} & \textcolor{brown}{5.5 $\mathrm{\pm}$ 0.5}\tabularnewline
\textcolor{brown}{500} & \textcolor{brown}{N$_{2}$} & \textcolor{brown}{0.1} & \textit{\textcolor{brown}{-}}\textcolor{brown}{5.8 $\mathrm{\pm}$
0.6} & \textcolor{brown}{8.6 $\mathrm{\pm}$ 0.9}\tabularnewline
\bottomrule
\end{tabular}
\end{table}

\begin{figure}
\noindent \begin{centering}
\includegraphics[width=8.5cm]{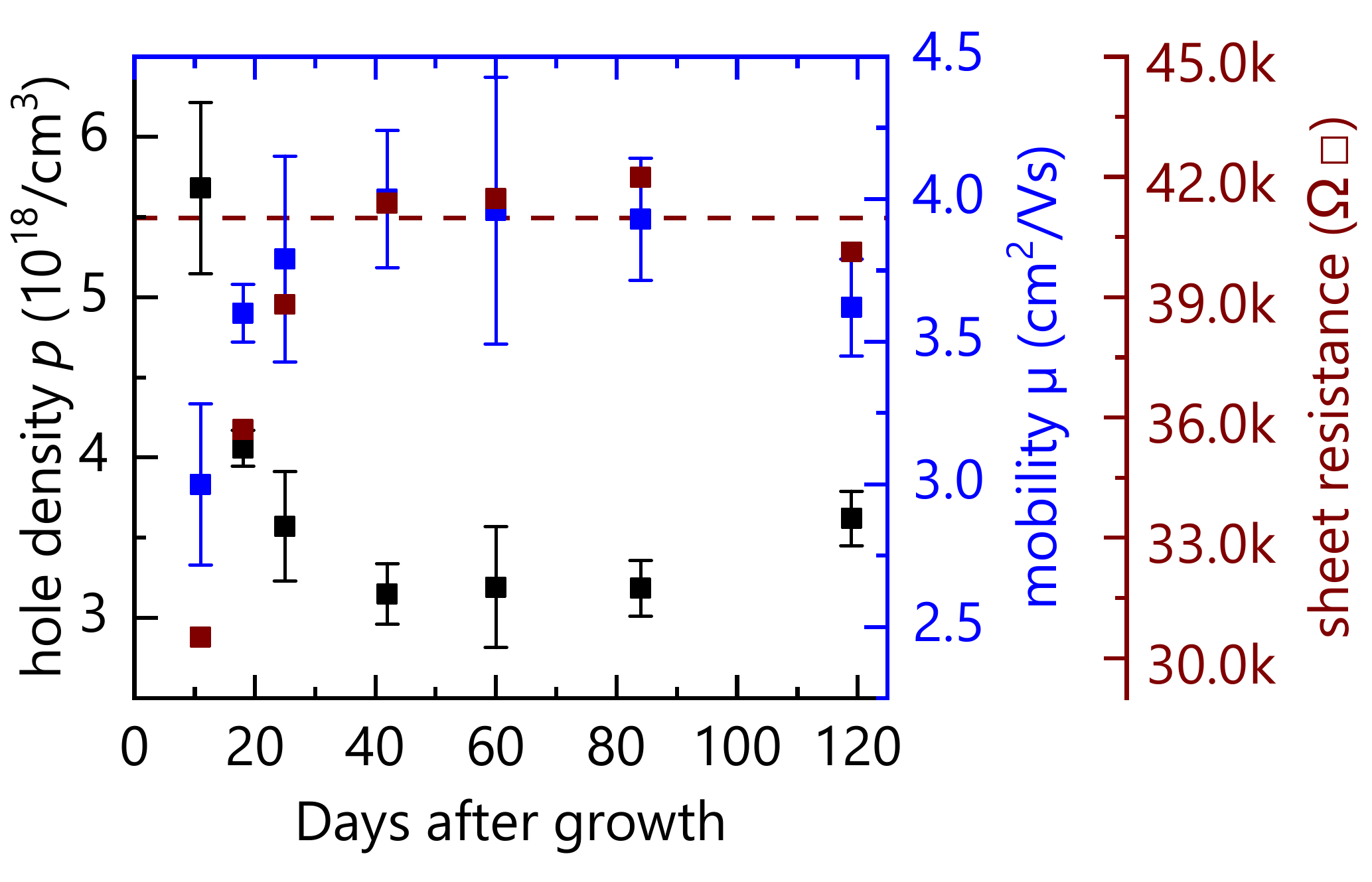}
\par\end{centering}
\caption{Hall hole concentration (black squares), sheet resistance (brown squares)
and Hall hole mobility (blue squares) of Y12 stored in ambient atmosphere
for up to 120 days after the growth run. The dashed brown line indicates
the saturation of the sheet resistance after 20 to 40 days. \label{fig:SnO_time stability}}
\end{figure}

\noindent In addition, the stability over time was investigated by
long\nobreakdash-term Hall measurements of Y12/a under storage in
ambient air. The measured Hall hole concentration, mobility and sheet
resistance are summarized in Fig.~\ref{fig:SnO_time stability} for
a period of 120 days. Only a slight change of the electrical properties
is found and a stabilization is indicated after about 40 days. No
change to \textit{n}\nobreakdash-type transport was observed.
 The long-term stability of the $p$-type transport in our SnO films
was confirmed in A400/a, A400{*}, and Y15/a that were re-measured
after a period of 14, 11, and 15 months, respectively.

\noindent These results identify an upper limit for the processing
or operating temperature between 300 and 400~$^{\circ}$C for devices using
SnO layers. Slight changes of the hole density with time or annealing
need further detailed investigation as they may relate to the unintentional
doping. The independency from the annealing atmosphere (with and without
O$_{2}$) suggests the change to \textit{n}\nobreakdash-type transport
upon annealing at 400\,$^{\circ}$C and above to occur by disproportionation
rather than oxidation. Thus, a capping layer would not allow processing
temperatures of $400$\,$^{\circ}$C and above. Long-term device operation
using SnO layers is feasable at room temperature in ambient air. According
to the equilibrium phase diagram (Fig.\,\ref{fig:phasediagram})
device operation should be feasable even up to 400\,$^{\circ}$C, which still
needs to be validated experimentally by long-term aging studies of
actual films at such temperatures.

\section{Concluding discussion}

Two key features of SnO are its metastability with respect to the
formation of (secondary) SnO$_{x}$ phases with $x=0$ or $1<x\le2$
as well as its anisotropic, comparably high hole mobility.

\subsection{Metastability}

The growth of the metastable $p$-type semiconducting oxide SnO is
challenged by its metastability, which we described by a theoretically
established\citep{factsage} phase diagram showing regions containing
Sn, SnO, and SnO$_{2}$. Adressing this metastability we presented
a rapid, experimental in-situ approach\citep{Vogt2015} to delineate
the SnO growth window in plasma-assisted molecular beam epitaxy based
on the understanding that SnO$_{2}$ grows through the intermediate
formation of its suboxide SnO.\citep{Vogt_twoStep} A critical assessment
of potentially present secondary phases was found to require the combination
of different experimental methods: While XRD is only sensitive to
epitaxially well-oriented phases (mainly SnO in our case), Raman spectroscopy
is essential, in particular for the identification of Sn$_{3}$O$_{4}$,\citep{Raman_Intermediate}
and SEM is key for the detection of Sn (droplets). Different from
the phase diagram, our experimental results point towards the formation
of Sn$_{3}$O$_{4}$ rather than SnO$_{2}$ as secondary phase for
slightly O-rich growth conditions. Near surface Sn$_{3}$O$_{4}$
detected by surface-sensitive Raman spectroscopy and a few-nm thick
surface SnO$_{2}$ layer detected by XPS are likely related to the
post-growth cooldown in oxygen plasma, and may possibly be avoided
by cooldown in vacuum. Layers with the closest stoichiometry to SnO
contain a fraction of either Sn$_{3}$O$_{4}$ or Sn as secondary
phase, necessitating a control of the Sn-to-O-plasma flux ratios better
than 10\,\% used in this study to achieve complete phase purity.
Room temperature hole transport properties and Seebeck coefficient
of these slightly O-rich and Sn-rich layers (samples Y12 and Y15)
are very similar, suggesting a minor impact of the secondary phases.
A significant enhancement of the hole mobility due to a secondary
Sn-phase reported in Ref.\,\citenum{Record_mob_SnO} was not reproduced
by our Sn-rich films (Y12 and Y10). Annealing experiments in different
atmospheres confirm structural and electrical stability of SnO layers
up to temperatures between 300 and 400\,$^{\circ}$C (in fair agreement with
the phase diagram), defining an upper limit for the thermal budget
of processing SnO-containing devices.

\subsection{Hole mobility}

At present the understanding of the hole mobility in SnO is rather
limited: The highest room temperature hole mobility in our (001)-oriented
films of 6.0\,cm$^{2}$/Vs at a hole concentration of $2.3\cdot10^{18}$\,cm$^{-3}$
in Y15/b was found to be mainly limited by phonon scattering in non-degenerate
band transport. The same holds true for the lower mobility of 2.5\,cm$^{2}$/Vs
in Y12/b at similar hole concentration. This behavior is, however,
inconsistent with the significantly higher theoretically predicted
Hall mobility limit due to phonons of $\mu_{\text{H}}^{\text{POP}}\approx13$\,cm$^{2}$/Vs
for this orientation,\citep{Hu2019} as well as the significantly
lower predicted Hall mobility for ionized impurity scattering of $\mu_{\text{H}}^{\text{IIS}}\approx2.3$\,cm$^{2}$/Vs
for an ionized acceptor concentration matching our hole concentration.\citep{Minohara2019}
Moreover, for the same orientation reports by Minohara et al.\citep{Minohara2019,Minohara2019a}
of $\mu_{\text{H}}\approx10$ to $21$\,cm$^{2}$/Vs at lower hole
concentration in single-crystalline layers document a degenerate behavior,\citep{Minohara2019a}
and hole transport limited by other mechanisms than phonon scattering.\citep{Minohara2019,Minohara2019a}
The contradiction of these results to our phonon-limited transport
at an even higher hole concentration suggests different types of transport,
possibly related to a significantly lower effective hole mass either
through a modified band structure or a defect band in the samples
by Minohara et al. An initial theoretical explanation involving the
effect of Sn interstitials and O vacancies on band structure has been
given by Granato et al..\citep{Granato2013} Other reported high hole
mobilities of $\mu_{\text{H}}\approx19$ and $30$\,cm$^{2}$/Vs
in the literature for polycrystalline material\citep{Record_mob_SnO,Miller2017}
may be related to transport through crystallites with the high-mobility
{[}001{]}-direction oriented in-plane, for which the theoretically
predicted Hall mobility limit due to phonons is $\mu_{\text{H}}^{\text{POP}}\approx106$\,cm$^{2}$/Vs.\citep{Hu2019}
While the anisotropy of the effective-mass and transport have been
experimentally determined and shown to match for the rutile $n$-type
oxide SnO$_{2}$,\citep{Button1971,Feneberg2014,Bierwagen2018} similar
experimental studies are missing for SnO to date.
\begin{acknowledgments}
\noindent We would like to thank H.-P. Sch\"onherr and C. Stemmler for
MBE support, A.-K. Bluhm for SEM imaging, S. Rauwerdink and A. Riedel
for sample processing, and M. Heilmann for critically reading the
manuscript. This work was performed in the framework of GraFOx, a
Leibniz-ScienceCampus partially funded by the Leibniz association.
M.B. and J.F. gratefully acknowledge financial support by the Leibniz
association. We are grateful to HiSOR, Hiroshima University, and JAEA/SPring-8
for the development of HAXPES at BL15XU of SPring-8. The HAXPES measurements
were performed under the approval of the NIMS Synchrotron X-ray Station
(Proposal No. 2018B4600, and 2019A4601).
\end{acknowledgments}

\bibliographystyle{apsrev4-1}
\bibliography{SnO}

\end{document}